\documentclass{aa}  

\usepackage{graphicx}
\usepackage{breqn}
\bibpunct{(}{)}{;}{a}{}{,}
\usepackage{natbib}
\usepackage{hyperref}
\usepackage{txfonts}
\usepackage{cancel}
\usepackage{pifont}
\usepackage{xcolor}
\usepackage{xspace}

\newcommand{\ekma}{\operatorname{{E\kern-.08em k}}} 
\newcommand{\nuss}{\operatorname{{N\kern-.09em u}}} 
\newcommand{\pram}{\operatorname{{P\kern-.03em m}}} 
\newcommand{\pran}{\operatorname{{P\kern-.03em r}}} 
\newcommand{\rayl}{\operatorname{{R\kern-.04em a}}} 
\newcommand{\reyn}{\operatorname{{R\kern-.04em e}}} 
\newcommand{\ross}{\operatorname{{R\kern-.04em o}}} 
\newcommand{\rosb}{\operatorname{{R\kern-.04em b}}} 
\newcommand{\stra}{\operatorname{{(N\kern-.04em / \Omega)^2}}} 
\newcommand{\straquatre}{\operatorname{{(N\kern-.04em / \Omega)^4}}}
\newcommand{\strasix}{\operatorname{{(N\kern-.04em / \Omega)^6}}}

\def \cesamxx{Cesam2k20\xspace}
\def \eos{EoS\xspace}

\begin{document} 

   \title{Magnetohydrodynamic instabilities in stellar radiative regions}
   \subtitle{I. Linear study of shear-driven instabilities}

   \author{V.~Durepaire\inst{1} \and L.~Petitdemange \inst{1} \and K.~Belkacem~\inst{1} \and  A.~Guseva\inst{2,1} \and L.~Manchon \inst{1} \and R.~Hollerbach \inst{3} \and F.~Daniel \inst{4}
          }

   \institute{LIRA, Observatoire de Paris, Université PSL, CNRS, Sorbonne Université, Université Paris Cité, 5 place Jules Janssen, 92195 Meudon, France\\
              \email{virgin.durepaire@obspm.fr}
             \and Department of Fluid Mechanics, Universitat Politècnica de Catalunya, BarcelonaTech (UPC), Barcelona 08019, Spain
             \and School of Mathematics, University of Leeds, Leeds LS2 9JT, UK 
             \and Center for Interdisciplinary Exploration and Research in
Astrophysics (CIERA), Northwestern University, Evanston, IL, USA
              }

   \date{Received December 12 2025; accepted January 31 2026}

 
   \abstract
   {Space missions such as Kepler have brought new constraints, along with new questions, on stellar evolution. A key issue is the unexpected spin-down of low-mass stellar cores, pointing to an efficient angular momentum transport mechanism in which magnetic fields are likely to play a role. This has renewed interest in the origin and impact of magnetic fields in stars.}
   {This paper is the first in a series investigating magnetohydrodynamic instabilities that might contribute to angular momentum transport and magnetic-field evolution in stellar radiative zones. Here, we focus on shear-driven instabilities and, specifically, the Goldreich-Schubert-Fricke (GSF) instability and the magnetorotational instability (MRI), which could play key roles in the internal dynamics of stellar radiative regions.}
   {We performed a detailed local linear stability analysis using a numerical approach that extends beyond classical limiting cases and incorporates stabilizing effects such as stratification and magnetic tension, enabling the exploration of more realistic flow regimes. These local results were then validated through a global mode analysis in a Taylor-Couette configuration. Together, these methods allow us to identify unstable regions, quantify growth rates, and assess the astrophysical relevance of the instabilities. Finally, we applied our results to evolutionary models of subgiant and young red giant stars constrained by recent observations.}
{We recovered the known standard MRI (SMRI) and azimuthal MRI stability criteria and quantified how stratification, magnetic tension, and diffusion affect their growth. In strongly sheared regimes, we derived a new criterion for the magnetised GSF (MGSF) instability and clarified how magnetic and stratification effects narrow the unstable domain, illuminating the transition from SMRI to MGSF. We also provided approximate growth time formulae that identify which instability (SMRI or MGSF) dominates under given stellar conditions and can be directly implemented in 1D stellar evolution codes to model angular momentum transport more realistically. Global Taylor-Couette calculations validate the local Wentzel-Kramers-Brillouin analysis, confirming that it is able to reliably predict unstable regions and mode behaviour.}
  {In its application to subgiants and young red giants, our results show that shear-driven instabilities can grow rapidly for magnetic fields below $100\,\mathrm{kG}$. The analytical criteria indicate where SMRI or MGSF modes should occur depending on the shear amplitude and location. Conversely, strong axial fields $(\sim 100\,\mathrm{kG})$ confined to the hydrogen-burning shell suppress instabilities unless the shear lies sufficiently far from the shell. These findings support incorporating our instability criteria and growth estimates into stellar evolution models to assess the efficiency of shear-driven transport.}

   \keywords{Magnetohydrodynamics (MHD) -- Plasmas -- Instabilities -- Stars: rotation -- Stars: evolution
               }
\authorrunning{V. Durepaire et al.}
\maketitle
%
\section{Introduction}
\label{sec:intro}
Solar-like oscillations in low-mass stars provide crucial information about their internal structure and dynamical processes. For example, mixed modes in evolved stars have revealed that red giant cores spin down along the red giant branch \citep{Mosser2012}, contrary to what would be expected under local conservation of angular momentum (AM). This discrepancy highlights our incomplete understanding of AM transport within stellar radiative regions, a long-standing problem in stellar physics. The issue is also unresolved at earlier evolutionary stages \citep[e.g.][]{Maeder2009, Fuller2019, Aerts2019} and one of the most striking cases is the nearly solid-body rotation of the Sun’s radiative zone \citep[e.g.][]{Garcia2007, Gough2015}. Several mechanisms have been proposed to account for the missing AM transport \citep[see][ for a review]{Aerts2019}, including transport associated with magnetic fields and their connection to internal dynamos \citep[e.g.][]{Eggenberger2019, Gouhier2022, Petitdemange2023}. However, no consensus has yet been reached thus far regarding the origin of the missing transport \citep[see][for a review]{Fuller2019}. Regarding the role of magnetic fields, not only does the associated transport remain uncertain, but, more broadly, the origin and strength of large-scale magnetic fields in stars are still a matter of debate \citep[e.g.][]{Braithwaite2017}. This is illustrated by the observed dichotomy between strong and weak magnetic fields in intermediate- and high-mass stars with radiative envelopes \citep[e.g.][]{Auriere2007, Lignieres2009}, as well as by the absence of detectable magnetic fields in rapidly rotating Be stars \citep{Wade2016}.

To advance our understanding of both the observed diversity of magnetic fields and the associated AM transport, magnetohydrodynamic instabilities (MHDIs) are a key physical ingredient that must be properly modelled within stellar radiative zones. Depending on the specific conditions, these instabilities can either destroy or sustain magnetic fields \citep{Lignieres2014, Braithwaite2017}, directly transport AM for instance through the magnetorotational instability (MRI) \citep{Balbus1991}, or redistribute AM indirectly through the turbulence or dynamo action that is generated, as in the case of the Tayler instability (TI) \citep{Spruit2002, Zahn2007}. 
In this context, numerical simulations have provided valuable insights into the non-linear behaviour of MHDIs and their efficiency in transporting AM \citep[e.g.][]{Barker2019, Guseva2017, Daniel2023, Petitdemange2023}. However, because they are dominated by viscosity, such simulations do not enable us to probe realistic stellar regimes. Consequently, scaling relations derived from numerical studies must be carefully extrapolated to stellar conditions \citep[e.g.][]{Ji2023}. Therefore, linear analyses remain essential, as they capture the onset of instabilities, provide general scaling relations, and guide prescriptions for 1D stellar evolution models. Indeed, simple prescriptions can be incorporated into 1D stellar evolution codes to account for the AM transport generated by MHDIs. Their impact is commonly represented through effective diffusivities \citep[e.g.][]{Cantiello2014, Wheeler2015, Griffiths2022}, making these parametrisations dependent on a solid theoretical foundation.

Among MHDIs, we can distinguish shear-driven instabilities (SDIs), which extract energy from differential rotation, from magnetically driven instabilities, which are powered by magnetic free energy. This first paper in the series focuses on SDIs and, in particular, the magnetorotational instability \citep[MRI,~][]{Balbus1991} as well as the Goldreich-Schubert-Fricke \citep[GSF,~][]{Goldreich1967, Fricke1968} instability. Our emphasis on the GSF instability is motivated by recent asteroseismic inferences suggesting the presence of strong shear in evolved stars \citep{Mosser2012, Deheuvels2014, DiMauro2016, Fellay2021}. Moreover, the GSF instability is often invoked as a source of angular momentum and chemical transport \citep[e.g.][]{Herwig2003, Barker2019, Chang2021} and is already implemented in some stellar evolution codes \citep[see][]{Paxton2013}. This underscores the need to clarify the conditions under which it would develop in the presence of a magnetic field, as well as to determine its associated properties.

Indeed, previous studies have shown that magnetic fields can stabilise GSF modes in certain regimes \citep{Caleo2016Insta, Caleo2016GSF, Dymott2024}. In this work, we extend these results to more general flow conditions, providing a new stability criterion and growth rate estimate that can be directly implemented in 1D stellar evolution models of magnetic stars, thereby allowing GSF-driven AM transport to explicitly account for magnetic stabilisation. At the same time, if a magnetic field is present, the flow might also become unstable to the MRI \citep{Velikhov1959, Balbus1991}; more precisely, to the standard MRI (SMRI) in the presence of a purely axial magnetic field, or to the azimuthal MRI (AMRI) when the field is predominantly azimuthal. In contrast with the GSF instability, the MRI is a weak-shear instability that requires a decreasing angular velocity outward, making it particularly relevant in regimes characterised by low differential rotation. While previous linear studies have identified thermal stratification as a stabilising factor \citep{Menou2004, Philidet2020}, we derived a general estimate to evaluate the efficiency of MRI development and show that the interplay between magnetic and thermal background fields can significantly slow its growth.

Local analyses provide a powerful tool for predicting MHDIs in stellar radiative regions, but they suffer from limitations. This is the case, in particular, in differentially rotating spherical geometries where the curvature, boundary conditions, and the finite size of the domain can inhibit or suppress their growth. To assess the limitations and the range of validity of our local approach, we  examine the stability of Taylor-Couette configurations with two differentially rotating cylinders, which offer a controlled framework for testing whether these modes can exist in practice \citep{Hollerbach2005, Guseva2015}. Such geometries have also been widely used to investigate AM transport and dynamo action driven by MHDIs \citep[][]{Guseva2017, Rudiger2025}. In this framework, the radial structure of the background flow is known to strongly influence the development of both MRI and GSF modes, providing valuable insight into the conditions required for these instabilities to operate in stellar radiative zones. More precisely, using this global approach, we show that the growth rates predicted by the local analysis provide a reliable indicator of the location of the SDI onset.
 
This paper is organised as follows. Section~\ref{sec:methodo} presents the local linear analysis and the underlying assumptions,  along with general estimates for MRI growth rates and onset conditions. Section~\ref{sec:GSF} describes the transition between the SMRI and GSF regimes, as well as the properties of the modified magnetised GSF. In Section~\ref{sec:global}, we determine the extent to which these local estimates can be applied to global linear instabilities using a magnetised Taylor-Couette model, which allows us to capture finite-domain effects. Section~\ref{sec:cesam} applies these results to subgiant and red-giant models and Section~\ref{sec:conclusion} summarises our main findings and perspectives.

\section{Local stability analysis and application to the MRI}
 \label{sec:methodo}

\subsection{Eigenvalue problem}

\begin{figure}
\centering
\includegraphics[width=7.5cm]{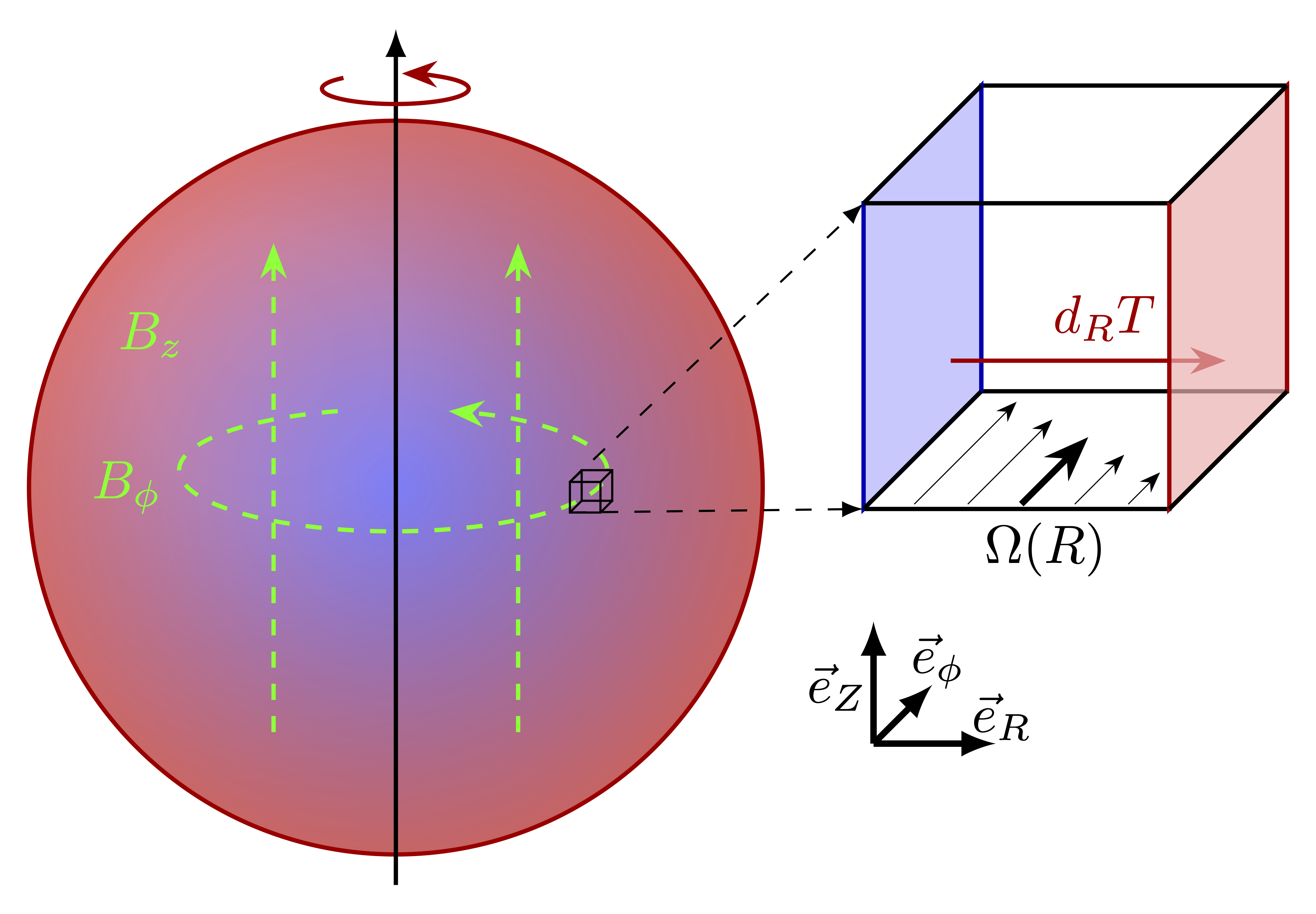} 
  \caption{Sketch of the system under study. In the inertial frame, the differentially rotating flow is confined between two shells, subjected to a radial thermal gradient and immersed in a magnetic field with both axial and azimuthal components.}
     \label{system}
\end{figure}

Motivated by previous studies \citep[e.g.][]{Acheson1978, Hollerbach2005}, we exploited the cylindrical symmetry of rapidly rotating flows and adopted cylindrical coordinates $(R,\phi,Z)$, which provide an appropriate local description near the equatorial region (see Fig.~\ref{system}). The analysis was carried out for a stably stratified regions within a static inertial frame, where we adopted the Boussinesq approximation. Therefore, the governing equations for momentum, induction, and temperature advection-diffusion, together with the divergence-free conditions, are expressed as:\ 
\begin{align}
&\vec \nabla \cdot \vec U = 0 \, , \quad \label{continuity} \vec \nabla \cdot \vec B = 0 \, , \\
&\dfrac{\partial \vec  U}{\partial t} + (\vec U \cdot \vec \nabla) \vec U = - \dfrac{\vec \nabla P}{\rho} + \frac{(\vec \nabla \times \vec B) \times \vec B}{\rho \mu_0} - \gamma T g \vec e_R+ \nu \vec \Delta \vec U \, , \label{NS} \\
&    \dfrac{ \partial \vec B}{\partial t} + (\vec U \cdot \vec \nabla) \vec B = (\vec B \cdot \vec \nabla) \vec U + \eta \vec \Delta \vec B \, , \label{Induction} \\
&\dfrac{\partial T}{\partial t} + (\vec U \cdot \vec \nabla) T = \kappa \Delta T \, . \label{Temper} 
\end{align}
Here, $\vec U$ is the velocity field, $\vec B$ the magnetic field,  $P$ is the pressure, $\rho$ the density, $T$ the temperature, $\gamma$ the coefficient of thermal expansion, $g$ the gravity, and $\nu$ and $\kappa$ are the kinematic viscosity and thermal diffusivity, respectively. 

Furthermore, Eqs.~(\ref{continuity}) to (\ref{Temper}) are perturbed around a background state. Each field is decomposed into a reference component and a perturbation, such that for a given quantity $\vec A,$  we obtain 
$\vec A = \vec A_0 \left(R,Z\right) + \vec A^\prime\left(R,\phi,Z,t\right)$. We assume that the reference state is characterised by a cylindrical rotation profile, so that $\vec U_0 = R\,  \Omega \left(R \right) \vec e_\phi$ and we decompose the magnetic field into axial and toroidal components $\vec B_0 = B_\phi \left(R \right)\Vec{e_\phi} + B_Z \, \vec{e_Z}$. The reference temperature profile $T_0(R)$ is assumed to be spherically symmetrical and the induced gravitational disturbances $g'$ are neglected. Within the Boussinesq framework, the buoyancy term is directly proportional to the temperature perturbation and the background temperature gradient is taken to be positive to represent a stably stratified medium. For the perturbations, we performed a local analysis at the equator, using the Wentzel-Kramers-Brillouin (WKB) approximation following \cite{Eckhoff1981} and \cite{Friedlander2003}. Hence, the perturbations were developed as\begin{align}
A^\prime = a \, \exp\left[\sigma t + i(k_R R + m \phi + k_Z Z)\right], \,  
\label{wave}
\end{align}
where $a$ represents the eigenmode’s relative amplitude in the linear regime, $m$ is the azimuthal degree, $k_R, k_Z$ are the radial and axial wavenumbers, $\sigma$ is the complex growth rate defined by $\sigma \equiv \sigma_r - i (\omega + m\Omega),$ with $\sigma_r$ the growth rate (if $\sigma_r >0$) or the decay rate (if $\sigma_r < 0$), $\omega$ is the frequency, and $m\Omega$ the frequency shift due to the relative motion of the perturbation (analogous to the Doppler effect). In line with the WKB approximation, the quantities associated with the background are considered uniform over the wavelength and curvature terms are neglected so that $k_R, k_Z \gg 1/R, m/R$ \footnote{Additionally, in radiative zones, density scale height $H_\rho = \vert \text{d} \ln \rho /\text{d}  R\vert ^{-1} \sim R$ (hydrostatic equilibrium gives $N^2=g/H_\rho$), so the Boussinesq condition $k_R H_\rho \gg 1$ is consistent with the WKB assumption. Similarly, one can define a shear scale $H_\Omega = \vert \text{d} \ln \Omega /\text{d}  R\vert ^{-1} = R/(2\vert \ross \vert)$, such that the WKB approximation requires $k_R R \gg 2 \vert \ross \vert$.}. In the following, the wavenumber and its components are scaled by $1/R$ and the complex growth rate, $\sigma$, is scaled by the local rotation frequency, $\Omega$, whilst keeping the same notation for the sake of brevity. The alignment of the mode with the rotation axis is quantified with $\alpha = k_Z/k$, where $k$ is the mode wavenumber. In addition,  the background gravity ($g_0$) is considered to be uniform and can be expressed more generally as the Brunt-Väissälä frequency, $\mathrm{N} = \sqrt{\gamma g_0 d_RT_0}$.

Using Eq.~\ref{wave} together with the short wavelength approximation for the perturbation into Eqs.~(\ref{NS}) to (\ref{Temper}) leads to the eigenvalue problem,  
\begin{align}
{\bf \mathcal{H}}\vec V^\prime = \sigma \, \vec V^\prime \, ,
\label{disp}
\end{align}
where we can use the solenoidal constraints in Eqs.~(\ref{continuity}) to reduce the system to five independent perturbations. The perturbation vector, $\vec V^\prime,$ is therefore defined as
\begin{align}
\vec V^\prime \equiv 
\begin{pmatrix}
    U^\prime_R & U^\prime_\phi & \dfrac{B^\prime_R}{\sqrt{\rho \mu_0}} & \dfrac{B^\prime_\phi}{\sqrt{\rho \mu_0}} &  \dfrac{\gamma g_0 T^\prime}{\Omega}
\end{pmatrix}^\top \, ,
\end{align}
while the matrix operator, $\mathcal{H,}$ is expressed as
\begin{align}
 &    \mathcal{H} \equiv  \\
 &    \begin{pmatrix}
    - \ekma k^2 & 2\alpha^2 & \mathcal{S}_B&  - 2\mathcal{T}_{B,\phi} \alpha^2  & \alpha^2 \\
    -2(1 + \ross) & -\ekma k^2  &  2\mathcal{T}_{B,\phi} (\rosb +1)   &  \mathcal{S}_B &  0 \\ 
     \mathcal{S}_B & 0 & -\ekma_\eta k^2   & 0 & 0 \\
    - 2\mathcal{T}_{B,\phi} \rosb &  \mathcal{S}_{B} & 2 \ross & -\ekma_\eta k^2  & 0 \\
    -\stra & 0 & 0 & 0 & -\ekma_\kappa k^2 \nonumber \\
\end{pmatrix},
\end{align}
where we adopted the notations as described in Table~\ref{table}. 

The diagonal terms in $\mathcal{H}$ correspond to viscous, magnetic, and thermal diffusions, expressed with $\ekma = \nu/\Omega R^2$ and  $\ekma_\eta  = \ekma / \pram,  ~ \ekma_\kappa  = \ekma /\pran$.  The terms $\mathcal{S}_{B}  = i (\mathcal{T}_{B,Z}+ \mathcal{T}_{ B,\phi})$, with $\mathcal{T}_{B,Z} = k_Z \sqrt{\Lambda_Z \ekma /\pram}$ and $ \mathcal{T}_{B,\phi} = m\sqrt{\Lambda_\phi \ekma /\pram}$, represent the magnetic stress coupling the velocity and magnetic perturbations. The angular velocity shear is defined as $\ross  \equiv (R/2\Omega)d_R \Omega$ and only regimes with $\ross <0$  are to be considered. Similarly, the magnetic shear is given by $\rosb \equiv (R^2/2B_\phi)d_R (B_\phi /R)$. The term $2 \ross$ represents the driving source of the magneto-rotational instability, while $-2(1+\ross)$ represents the source of the hydrodynamic shear instabilities. To avoid any magnetic kink instabilities, our focus was exclusively on the stabilizing effect of magnetic pressure and set $\rosb = -1$. We also note that in the limiting isothermal case where buoyancy effects vanish and the GSF cannot develop, the formulation was reduced to that of \cite{Kirillov2014}. 

To maintain physical consistency, we imposed a geometric constraint on the admissible modes, such that their wavelengths needed to remain smaller than the characteristic system size (i.e. $k_R R \geq 2 \pi$ and $k_ZR \geq  2 \pi$). This condition is slightly more permissive than the strict requirement for the validity of the WKB approximation. Furthermore, to exclude modes that are strongly damped by viscosity, we required the viscous damping rate to remain smaller than the typical growth rate (i.e. $\nu k_{R}^2, \nu k_{Z}^2 \leq \vert \ross \vert \Omega; $ see Sect.~\ref{sec:growthrateMRI}). 
With the numerical method described in Appendix~\ref{num_met}, we then solve the eigenvalue problem given in Eq. ~\ref{disp} for a set of wavevector. We then retained the mode with the largest growth rate, as it is expected to govern the onset of the instability. This method only allows us to study low azimuthal wavenumber $(m \ll k_R R, k_Z R)$. Our results can be viewed as indicative of the trends one might expect when extending the analysis to larger values of $m$. However, such an extrapolation implicitly assumes that curvature and other $m$-dependent effects remain negligible. 

\subsection{Magneto-rotational instabilities (MRI)}
\label{sec:MRI}
\subsubsection{Validation of the local approach}

\begin{figure}
\centering
\includegraphics[width=8.6cm]{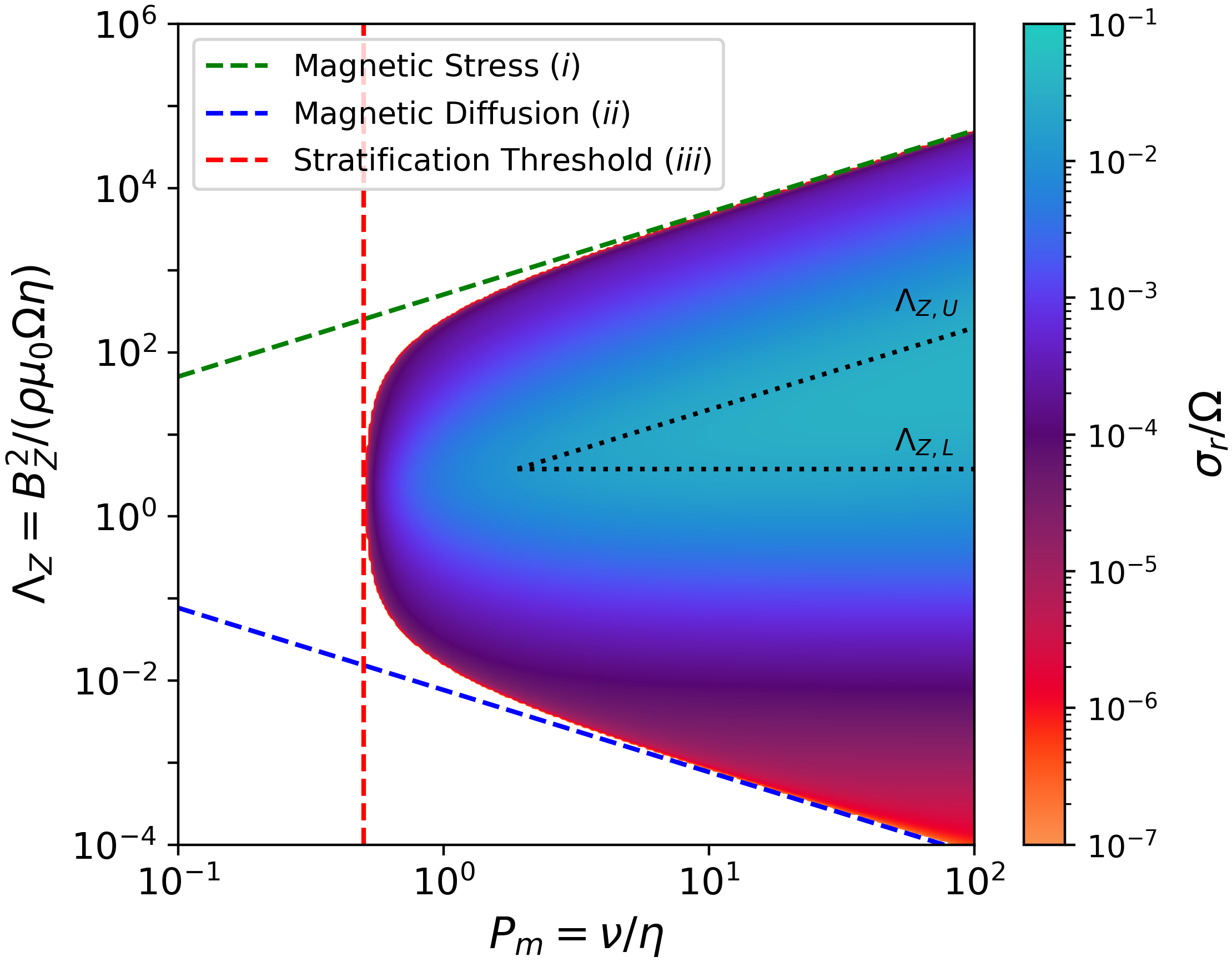} \\
\includegraphics[width=8.6cm]{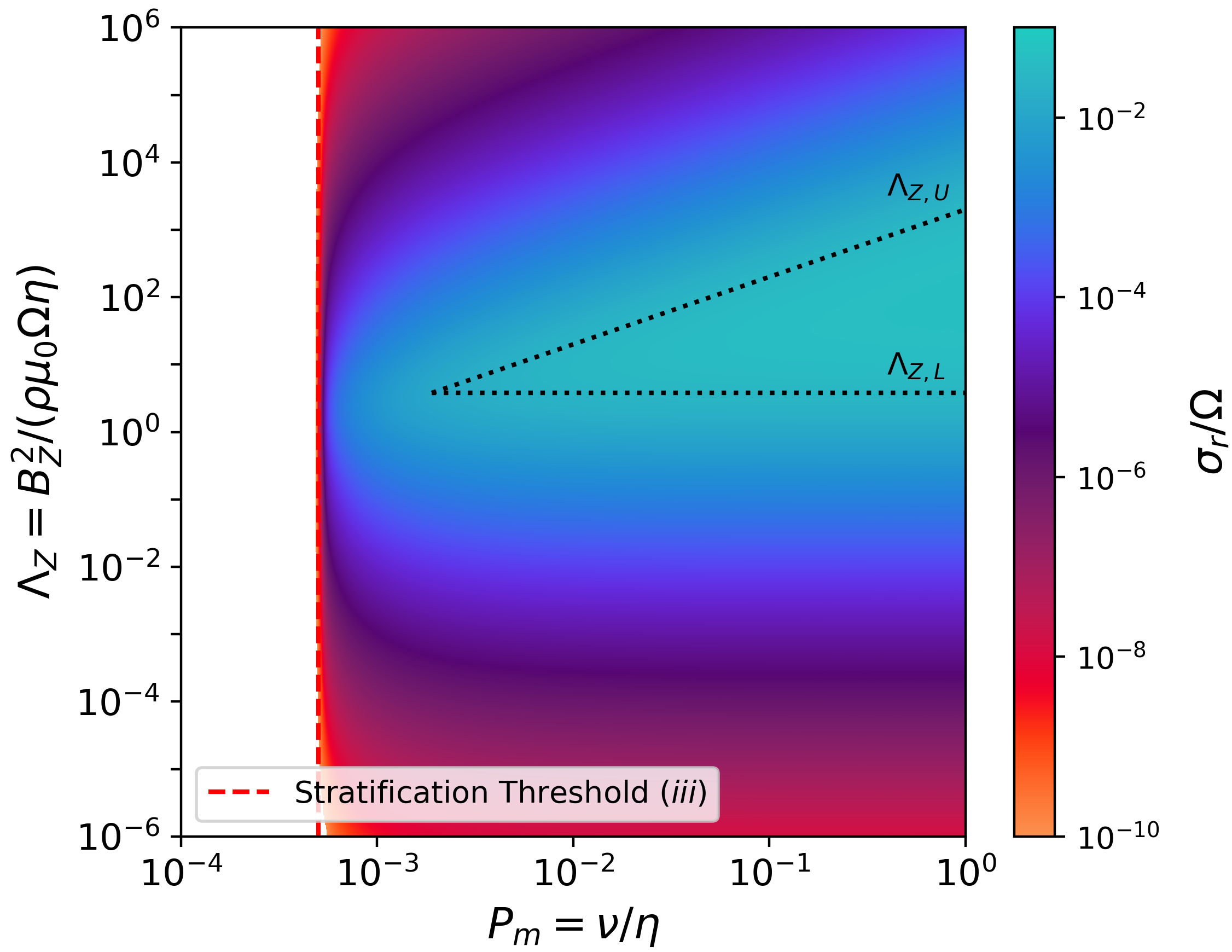}
  \caption{
  \textbf{Top panel:} 
  Growth rate, $\sigma_r$, normalised by $\Omega$ in the $(\pram, \Lambda_Z)$ plane of the axisymmetric SMRI, for a regime typical of DNS. See details in Table~\ref{table}, with $\ross = -0.05, \stra = 1, \Lambda_\phi = 0$. The three dashed lines correspond respectively to the three criteria equation \ref{MRI_crit}. The blank region corresponds to the stable parameter domain. The dotted lines correspond to $\Lambda_{Z,L}, \Lambda_{Z,U}$, from the Eq.~\ref{bandpass}, when $\Lambda_{Z,U} > \Lambda_{Z,L}$.
  \textbf{Bottom panel:} Growth rate, $\sigma_r$, normalised by $\Omega$ in the $(\pram, \Lambda_Z)$ for a regime of radiative stellar region (see Table~\ref{table}, with $
  \ross = -0.05, \stra = 10, \Lambda_\phi = 0$). 
          }
     \label{expli}
\end{figure}

\begin{table}
\centering
\renewcommand{\arraystretch}{1.3}
\caption{Dimensionless parameter set (Prandtl, magnetic Prandtl, Ekman, Elsässer, azimuthal Elsässer, stratification, and shear-rate number), with typical values used in DNS and estimates for stellar radiative regions.}
\begin{tabular}{lll}
\hline\hline
     \textbf{Parameter} & \textbf{DNS} & \textbf{Stellar rad. reg.} \\
     \hline
     $\pran = \nu / \kappa$ & $10^{-1}$ & $10^{-5}$ \\
     $\pram = \nu / \eta$ & $1$ & $10^{-3}$ \\
     $\ekma  = \nu / (\Omega R^2)$ & $10^{-5}$ & $10^{-15}$ \\
    $\Lambda_{Z} = B_Z^2/(\rho \mu_0 \Omega \eta) $ & $1$ & $10^{-5}-10^{15}$ \\
    $\Lambda_{\phi} = B_\phi^2/(\rho \mu_0 \Omega \eta) $ & $1$ & $10^{-5}-10^{15}$ \\
    $\stra = \gamma g_0 d_R T_0/\Omega^2$ & $1$ & $1 - 10^6$ \\
    $\vert \ross \vert = \vert (R/2\Omega)d_R \Omega \vert $ & $10^{-2} - 1$ & $10^{-5}-100$ \\
    \hline
    \end{tabular}
\tablefoot{For the DNS regimes see for instance the parameters used in \citet{Petitdemange2023}. For the stellar radiative regions, based on the theoretical analyses of \citet{Zahn1992, Spruit2002, Mathis2005}, and summarised by \citet{Maeder2009}, order of magnitudes of flow properties. The magnetic strengths considered account for the large variety of magnetic fields observed \citep{Reiners2012}. }
    \label{table}
\end{table}
We present here the SMRI parameter domains using the same representation as \citet{Philidet2020}. To do so, we reformulated the eigenvalue problem in terms of adapted dimensionless parameters (see Table~\ref{table}). The key numbers considered are the Ekman, axial Elsässer, Prandtl, magnetic Prandtl, stratification, and shear numbers, denoted  $(\ekma, \Lambda_Z, \pran, \pram, \stra, \ross)$. For the SMRI, non-axisymmetric perturbations differ from the axisymmetric case only by a frequency shift of $+m\Omega$, so the study is restricted to $m=0$. Figure~\ref{expli} shows the phase diagram of system stability in the $(\pram,\Lambda_Z)$ plane for parameter values typical of DNS and stellar radiative regions. 

The main limits within the phase diagram are well-known \citep[see][]{Chandra1960, Acheson1978, Balbus1991}: below a given $\pram_{\mathrm{min}}$, the instability does not develop. Above that critical value one can also see that the system is only unstable if $\Lambda_Z$ is in a certain window, whose width increases with $\pram$. The upper bound in terms of $\Lambda_Z$ corresponds to an excessive magnetic strength. The lower bound in terms of $\Lambda_Z$ is imposed by magnetic diffusion. In addition, as shown by \citet{Menou2004} and \citet{Philidet2020}, stratification has a stabilizing influence on SMRI modes, setting the threshold value $\pram_{\mathrm{min}}$ that depends on the competition between shear and stratification. 
Combining these representations, we recovered the three well-known criteria, namely,

\begin{equation}
\begin{aligned}
\begin{cases}
 4\dfrac{k_Z^2}{k^2} \pram \left\lvert \ross \right\rvert < \left(k_Z^2 \Lambda_Z + m^2 \Lambda_\phi \right) \ekma,
&\hspace{1em} \textbf{\textit{(i)}} \\[2ex]
\left\lvert \ross \right\rvert \left[1+ \left(k_Z^2 \Lambda_Z + m^2 \Lambda_\phi \right) \dfrac{\pram}{\ekma k^4} \right]< 1+ \dfrac{\pran}{4}\stra, &\hspace{1em} \textbf{\textit{(ii)}}          \\[2ex]
4 \pram \left\lvert \ross \right\rvert < \pran \stra.
&\hspace{1em} \textbf{\textit{(iii)}}
\end{cases}
\end{aligned}
\label{MRI_crit}
\end{equation}

The limits obtained with our numerical method are in very good agreement with the theoretical criteria. As can be seen by comparing the two panels of  Fig.~\ref{expli}, the range of $\Lambda_Z$ values that permits the development of the SMRI is much broader in realistic stellar regimes than in typical DNS regimes. This reflects the fact that diffusive processes are comparatively inefficient in radiative regions, so the window of unstable modes is wider. In  Appendix~\ref{ApMRI}, we display the stability domain of the same regime parameters towards AMRI\footnote{It requires $m \neq 0$ in order to generate a finite azimuthal magnetic tension $(mB_\phi/R)$ that drives the instability. The case $B_\phi \neq 0$ with $m=0$ does not excite the AMRI but can instead give rise to pinch or kink-type instabilities from the magnetic gradient $d_R B_\phi$, which are of different nature and outside the scope of this work.}, where the azimuthal Elsässer $\Lambda_\phi = B_{\phi}^2/(\rho \mu_0 \Omega \eta)$ has to be considered instead of $\Lambda_Z$. The limit of the unstable parameter domain corresponds also to the limits of Eq.~\ref{MRI_crit}.
 
\subsubsection{Growth rates of the MRI}
\label{sec:growthrateMRI}

The purpose of this sub-section is to derive general expressions for the growth rates of both SMRI and AMRI to determine the conditions under which an effective MRI can operate. Although these approximate expressions only hold when the perturbations remain small compared to the background quantities, they are useful for estimating which type of instability is likely to dominate and develop efficiently. While non-linear saturations may modify the general dynamics, the expressions we develop next are of interest for implementation in stellar evolution codes since an instability must have a growth time ($\tau_r = \sigma_r^{-1}$) shorter than the typical stellar evolutionary timescale ($\tau_{\mathrm{evol}}$) to be effective.

Figure \ref{expli} shows the growth rate of the SMRI. Our calculation exhibits a plateau within which the instability behaves ideally and approaches its maximum growth rate ($\sigma_{0}$), which is equal to $\vert \ross \vert \, \Omega$ for $\vert \ross \vert \le 2$ and to $2 \sqrt{\vert \ross \vert - 1} \, \Omega$ otherwise. This plateau is bounded by lower and upper limits on the magnetic field strength, 

\begin{equation}
\Lambda_{Z,L} \equiv 4\vert 1+\ross\vert, \qquad \Lambda_{Z,U} \equiv \left\vert \ross / \ross_{\mathrm{min}} \right\vert,
\end{equation}
\noindent
with $\ross_{\mathrm{min}}$ defined by Eq.~\ref{bandpass}. These limits are displayed in Fig.~\ref{expli} and correspond, respectively, to the magnetic field strengths at which magnetic stress balances either diffusion combined with the Coriolis effect, or magnetic tension competes with stratification to counteract shear. The first limit is consistent with the non-stratified case derived by \citet{Petitdemange2008}, while the second extends this framework by identifying a new stabilizing regime arising from the interplay between magnetic tension and buoyancy. 

To go further, because this growth rate can serve both to identify the dominant instability in simulations and to estimate the associated angular momentum transport in stellar evolution codes, we propose a simple analytical fit that provides accurate values without solving the full eigenvalue problem across the unstable parameter domain (Eq.~\ref{disp}). Guided by the numerical growth rates together with Eq.~\ref{MRI_crit}, we found that the following expression best reproduces our calculation,\emph{}
\begin{equation}
\begin{aligned}
\sigma_{\mathrm{SMRI}} \equiv \sigma_0 \left[1+Q^2\left(\dfrac{\Lambda_Z}{\Lambda_{Z,U}} + \dfrac{\Lambda_{Z,L}}{\Lambda_Z}\right) \right]^{-1},  
\label{gr_model}
\end{aligned}
\end{equation}
with
\begin{equation}
\begin{aligned}
Q^2 &= \left(1-\left \vert \dfrac{\ross_{\mathrm{min}}}{\ross}\right \vert \right)^{-2},
\qquad
\vert \ross_{\mathrm{min}} \vert~ = \dfrac{\pran \stra}{4 \pram}.
\end{aligned}
\label{bandpass}
\end{equation}
For the regimes shown in Fig.~\ref{expli} (top and bottom panels), the approximated formula (Eq.~\ref{gr_model}) is consistently close to the numerical values of $\sigma_{\mathrm{r}}$, lying between $\sigma_{\mathrm{r}}$ and $1.46 \times \sigma_{\mathrm{r}}$. Across all other regimes tested, $\sigma_{\mathrm{SMRI}}$ reproduces the numerical growth rates $\sigma_{\mathrm{r}}$ fairly well. We also note that for a given flow regime, the maximum growth rate occurs for a magnetic field strength $\Lambda_{Z,C}$ such that $\Lambda_{Z,C}^2 = \Lambda_{Z,L} \times \Lambda_{Z,U}$, as implied by Eq.~\ref{gr_model}.

For the AMRI (as detailed in Appendix~\ref{ApMRI}) the squared growth rate scales as $\Lambda_\phi/\pram$ for a given $m$. Consequently, unlike the SMRI, they do not reach a plateau over a given range of magnetic field strengths. However, as for the SMRI, it is possible to reproduce the numerical growth rates and it is expressed as
\begin{equation}
\begin{aligned}
       \sigma_{\mathrm{AMRI}} \equiv \sqrt{\dfrac{m^2 \Lambda_\phi \ekma \vert \ross \vert}{\pram}} \left( 1-  \dfrac{\Lambda_{\phi,L}}{\Lambda_\phi }  \right) \left( 1-  \dfrac{\Lambda_{\phi}}{\Lambda_{\phi,U}}  - \left \vert \dfrac{\ross_{\mathrm{min}}}{\ross}\right \vert \right)^{2},
    \label{grAMRI2}
\end{aligned}
\end{equation}
with 
\begin{equation}
\Lambda_{\phi,L} = \ekma \dfrac{k_{min}^4}{m^2} \left( \dfrac{1}{\vert \ross \vert} + \dfrac{\pran \stra}{4 \vert \ross \vert } -1 \right), ~\Lambda_{\phi,U} = 4\dfrac{\pram \vert \ross \vert}{\ekma m^2} \, .
\end{equation}
Equation~(\ref{grAMRI2}) is consistently close to the numerical result with $\sigma_{\mathrm{r}}$ ranging between $0.3 \sigma_{\mathrm{r}}$ and $3 \sigma_{\mathrm{r}}$. It deviates significantly from $\sigma_{\mathrm{r}}$ only near the boundary of the unstable domain. The simple expression $\sigma_{\mathrm{AMRI}}$ further reproduces rather well the numerical values $\sigma_{\mathrm{r}}$ across all other regimes tested. Its worth noting that both Eq.~(\ref{grAMRI2}) and numerical results reach their maximum for $\Lambda_{\phi, U} / 5$ and when $\vert \ross\vert \gg \ross_{\mathrm{min}}$ one has $\sigma_{\mathrm{AMRI}} (\Lambda_{\phi,C}) \sim \sigma_0$. 

Since both $\Lambda_{\phi,L}$ and $\Lambda_{\phi,U}$ depend on $m$, unstable azimuthal magnetic fields satisfy $\Lambda_\phi < \Lambda_{\phi,U}(m=1)$. Moreover, if an azimuthal wavenumber, $m$, exists such that $m \ll k_R R, k_Z R$ and $\Lambda_\phi \sim \Lambda_{\phi, U}/5$, the characteristic growth time approaches the ideal timescale, $\tau_r \sim \tau_0$. This result agrees with the local prediction from the local linear theory of \citet{Masada2006}. For weaker magnetic fields, high $m$ values that fall outside the strict validity of the local approximation are required; this point is discussed in Sect.~\ref{sec:global}.

\begin{figure}
\centering
\includegraphics[width=8.6cm]{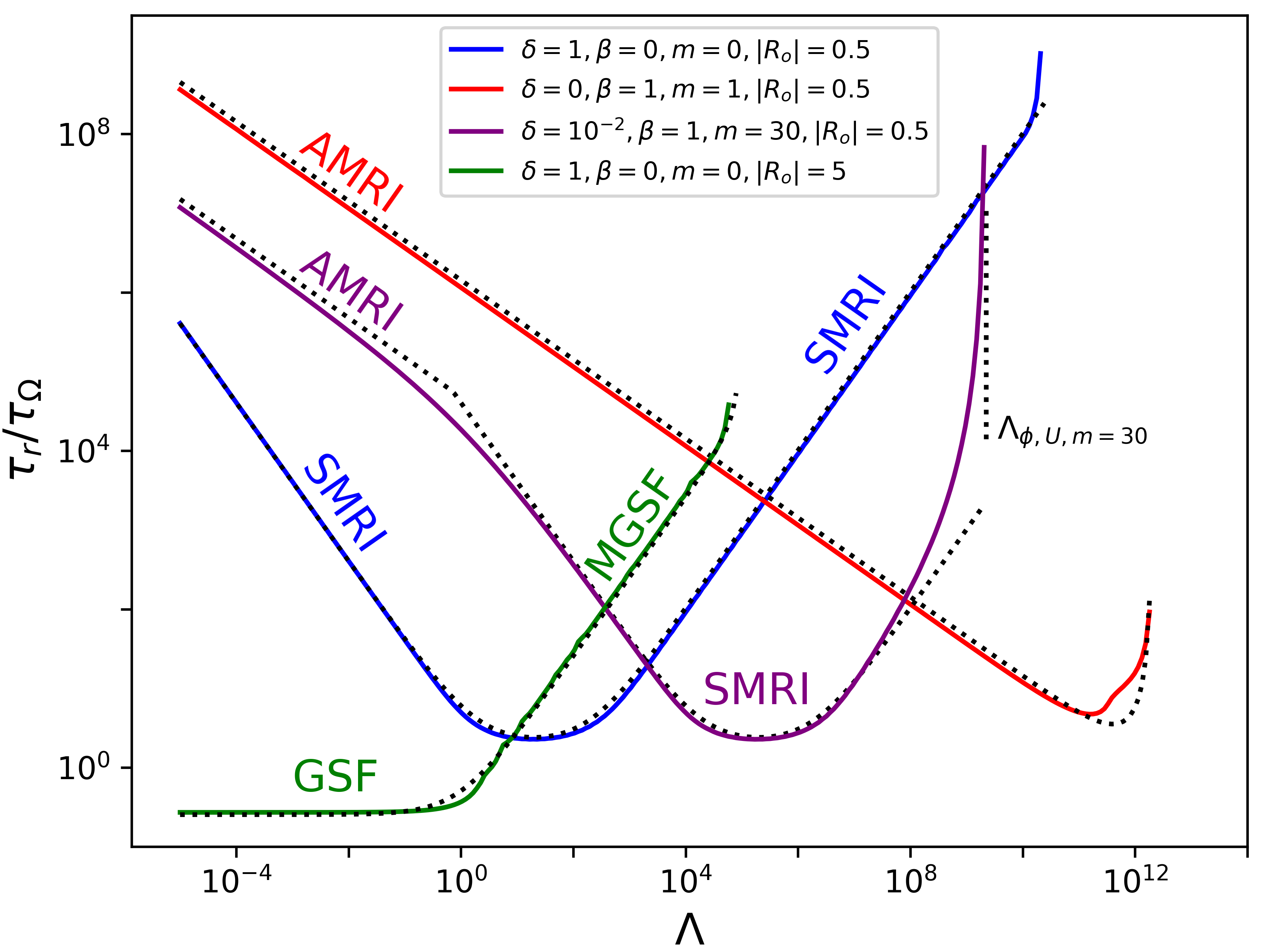}
  \caption{Growth time, $\tau_r/\tau_\Omega (=\Omega/\sigma_r),$ as a function of  $\Lambda$ for different magnetic-field configurations, parametrised by $\Lambda_Z = \delta^2 \Lambda$, $\Lambda_\phi = \beta^2 \Lambda$. We consider regimes representative of stellar radiative zones (Table~\ref{table}): a weakly stratified, low-shear case $(\vert \ross \vert = 0.5, \stra = 1)$ relevant to the MRI, and a strongly stratified, high-shear case $(\vert \ross \vert = 5, \stra = 5\times10^3)$ relevant to the GSF and MGSF instability (Sect.~\ref{sec:GSF}).
The dotted lines show the analytical growth rate estimates (Eqs.~\ref{gr_model}, \ref{grAMRI2}, and \ref{estimMGSF}).
  }
     \label{GR}
\end{figure}

Figure~\ref{GR} shows that the approximated expressions of the growth rate (Eqs.~\ref{gr_model} and \ref{grAMRI2}) reliably reproduce the instability timescales across different magnetic configurations. With respect to the helical magnetic field, for $\Lambda_Z < \Lambda_{Z,C}$, the SMRI and AMRI form distinct branches and the fastest mode dominates; when they become simultaneously competitive, a reinforced hybrid mode may occur. For stronger axial fields ($\Lambda_Z > \Lambda_{Z,C}$), the AMRI is suppressed and only the SMRI remains, while sufficiently large azimuthal magnetic stress ($\Lambda_{\phi} > \Lambda_{\phi,\mathrm{max}}$) stabilises the SMRI itself. Overall, our findings show an excellent agreement with the linear analysis of SMRI and AMRI reported in \citet{Petitdemange2013} and further extend this framework by incorporating stabilising thermal stratification. Finally, we note that in the parameter regimes explored here, we did not recover the helical MRI branch described by \citet{Liu2006} and \citet{Kirillov2014} \footnote{Note that we reproduce \citet{Kirillov2014}.}. Instead, we found only a helically modified MRI, which differs only slightly from the SMRI or AMRI. For this reason, we did not pursue the study of the helical MRI further in the context of radiative stellar interiors. 
In Appendix~\ref{ApMRI}, we provide additional details on the mode structure of the SMRI and AMRI and give local prescriptions for the radial and vertical scales of the unstable modes. 

\section{Magnetised GSF (MGSF) instability}
 \label{sec:GSF}

\subsection{Transition from SMRI to GSF}

We first considered the transition from SMRI to GSF in flows where the magnetic field is too weak to significantly modify the unstable domain. This allows us to identify how magnetic fields alter the properties of the instabilities and how, together with the effect of stratification, SMRI modes can smoothly transition into magnetised GSF (MGSF) modes.

\begin{figure}
\centering
\includegraphics[width=8.6cm]{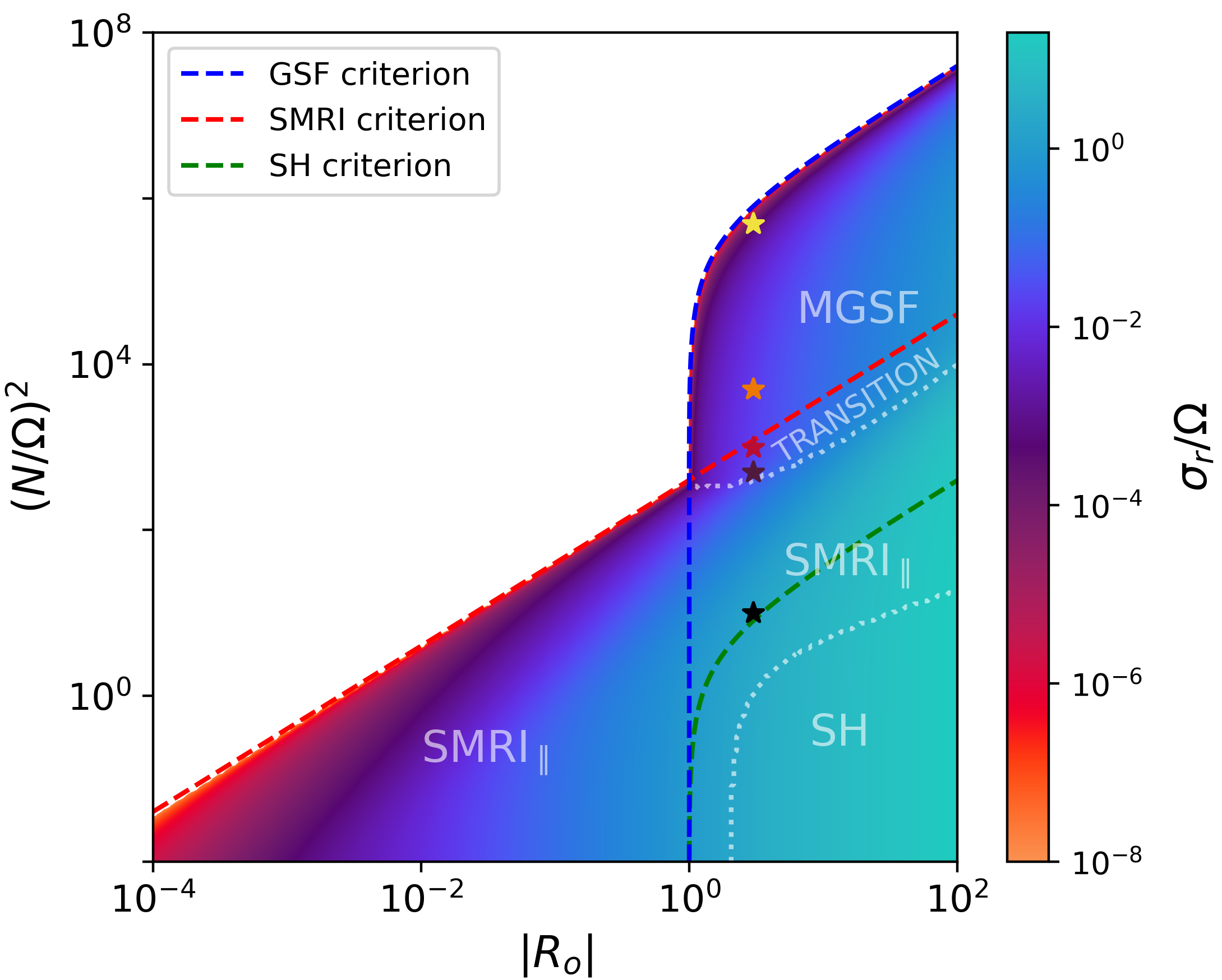}
\includegraphics[width=8.6cm]{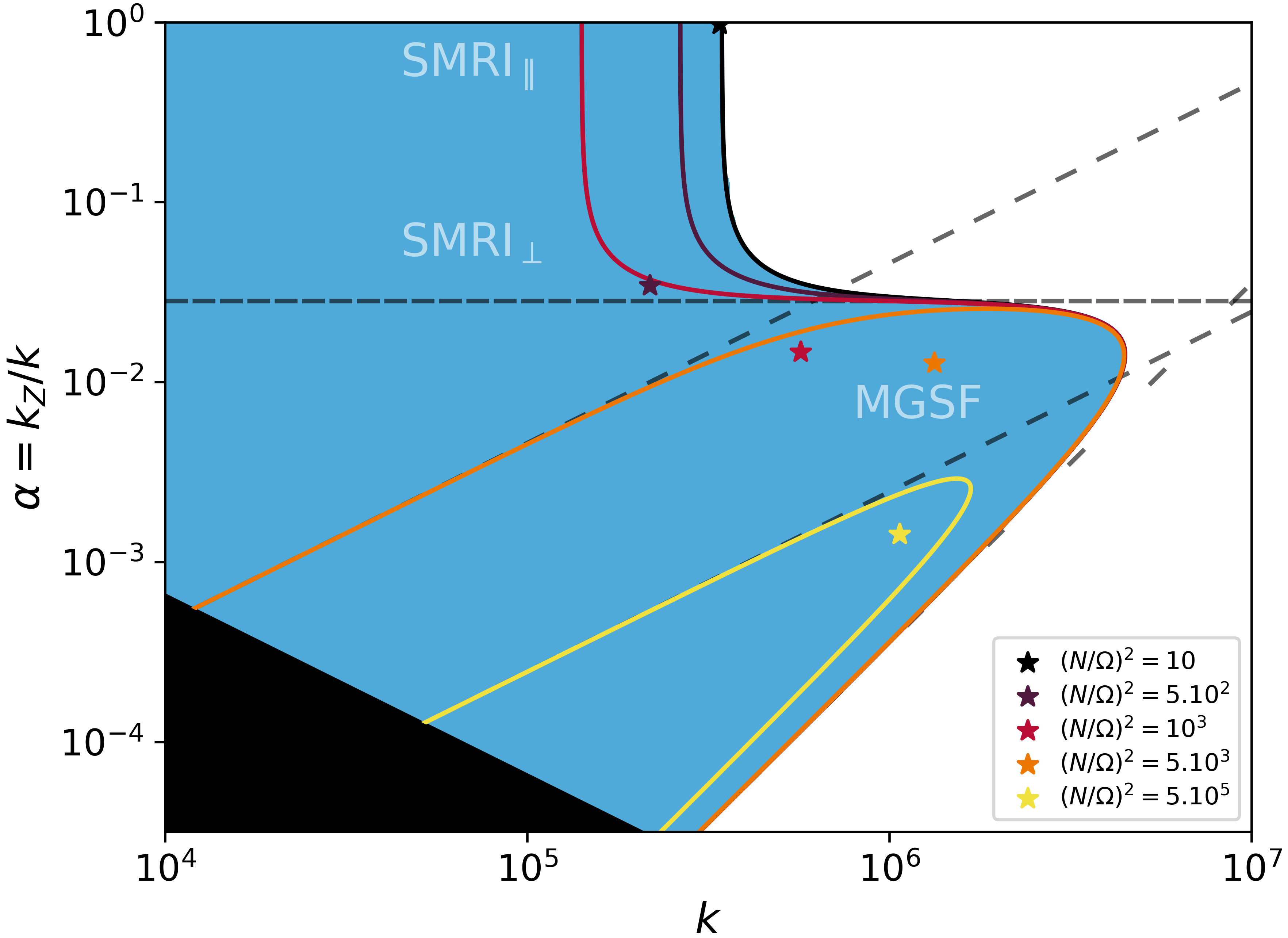}
  \caption{\textbf{Top panel:} Growth rate $\sigma_r / \Omega$ in the $(\vert \ross \vert, \stra)$ of the SMRI, GSF and SH instabilities, for a regime typical of radiative regions (see Table~\ref{table}, $\Lambda_Z = 100, \Lambda_\phi = 0 $). The three dashed lines correspond to instability thresholds, red for SMRI (Eq.~\ref{MRI_crit}), blue for GSF, green for SH. The dotted lines separate the dominant type of instability inside the unstable domain from the mode properties. \textbf{Bottom Panel:} Stability map in the $(k,\alpha)$ plane for a stellar regime with $\lvert \ross \rvert = 3$ and $\Lambda_Z = 100$. Solid colored curves show the boundaries of the unstable domains for several values of the stratification parameter $\stra$. The blue region corresponds to the unstable range for $\stra = 10$, and the star markers indicate, for each $\stra$, the most unstable mode (corresponding to the regimes indicated by the stars in the top panel). The black region denotes modes excluded by our assumptions. The densely dashed horizontal line indicates the upper limit of the GSF-unstable domain set by magnetic stresses. The dashed lines show the modified stability boundaries due to the combined action of magnetic stresses and stratification, shown here for $\stra = 5\times 10^{3}$ and $5\times 10^{5}$. The loosely dashed line marks the limit imposed by viscous diffusion. Labels indicate the domains of SMRI\(_\parallel\), SMRI\(_\perp\), and MGSF modes.
 }
     \label{sSMRI}
\end{figure}

In Fig.~\ref{sSMRI} (top panel), the phase diagram $( \vert \ross \vert, \stra )$ is displayed for a sheared flow in a regime typical of radiative regions
(see Table~\ref{table}) with $\Lambda_Z = 100$. The dashed lines show the boundary between the stable and unstable domains for each instability\footnote{We recall that the standard GSF and SH criteria read $4\left(\vert \ross \vert-1\right) >\pran \stra$ and $4\left(\vert \ross \vert-1\right) > \stra$.}. The SH modes correspond to large scales (small $k$), SMRI to intermediate scales, and MGSF to smaller scales. For sufficiently strong magnetic fields, SMRI modes (SMRI$_\parallel$) remain aligned with the field, whereas MGSF modes become nearly perpendicular, reflecting the balance between shear destabilisation and magnetic stabilisation. As shown in Fig.~\ref{sSMRI} (bottom panel), weak stratification produces a broad SMRI-dominated unstable range, while stronger stratification narrows this range and shifts the orientation of the most unstable mode. Since the transition from SMRI to MGSF is continuous in the $(k, k_Z/k)$ plane, the modes evolve smoothly between the two, passing through SMRI-like modes that become nearly perpendicular to the rotation and magnetic-field axes (SMRI$_\perp$). This gradual re-orientation resembles the behaviour reported by \citet{Dymott2024}, who showed that magnetic fields tilt the GSF-unstable domain and produce SMRI modes aligned with the angular momentum gradient. 

Once stratification stabilises the SMRI, the MGSF domain is limited by resistive effects that impose an upper limit. In addition, the magnetic stress restrict the mode orientation, the viscous effects damp small-scale modes, and the magneto-stratification excludes large-scale modes. These properties of the MGSF domain and mode properties are quantified and detailed in Appendix~\ref{sec:marginal}.

\subsection{Magnetised GSF:\ Stability criterion}

While some stellar models appear compatible with the development of GSF modes \citep[e.g.][]{Fellay2021}, recent studies have shown that magnetic fields can strongly damp these modes \citep{Caleo2016GSF, Dymott2024}. Because these results were obtained in specific parameter regimes, our goal is now to derive general stability criteria that explicitly account for the stabilizing effect of an axial magnetic field, thereby allowing us to assess the occurrence of GSF modes in a more realistic stellar context.

To this end, following the approach used to determine the unstable mode range, we used the marginal stability equation to examine two regimes. Compared with the standard criterion, we incorporated additional stabilizing forces and by focussing on the most unstable mode, we were able to derive new conditions. We first considered a regime in which the additional stabilizing forces are due to mixed diffusion and magnetic stress. By identifying the mode $(k, k_Z/k)$ that minimises stabilisation, we obtained a first correction term (hereafter, $\mathcal{C}_1$, defined in Eq.~\ref{def_coeffs_c}) to the stability criterion. Second, we add the stabilizing forces due to viscous diffusion and the magneto-stratification interplay. In this case, in a similar way, we obtained a second correction term (hereafter, $\mathcal{C}_2$, defined in Eq.~\ref{def_coeffs_c}). The detailed derivation (described in Appendix~\ref{ApMGSF}) finally gives a sufficient MGSF stability criterion of
\begin{equation}
\begin{aligned}
\pran \stra &+ \mathcal{C}_1 + \mathcal{C}_2 > 4\left( \vert \ross \vert-1 \right), 
\end{aligned}
\label{MGSFcrit}
\end{equation}

where 
\begin{equation}
\begin{aligned}
\label{def_coeffs_c}
\mathcal{C}_1 &= 2 \sqrt{2}  k_{Z,min} \Lambda_Z^{3/2} \ekma ^{1/2}, \\
\mathcal{C}_2 &= \dfrac{5}{2}\left(\dfrac{2}{3}\right)^{3/5} \left(k_{Z,min}^2\ekma \Lambda_Z^3 \pran^3 \left[\stra - 4 \pram \vert \ross \vert /\pran\right]^3 \right)^{1/5}. 
\end{aligned}
\end{equation}
The mode $k_{Z,min}$ designates the mode with the largest axial extent allowed within the domain and, therefore, the non-dimensional vertical wavenumber is expressed as $k_{Z,min}=2 \pi$. In Fig~\ref{GSFcri}, the phase diagram $ (\vert \ross \vert, \stra)$ of a stellar regime with a strong magnetic field ($\Lambda_Z=10^5$) is displayed. The limits of the unstable domain are well reproduced by the criteria of the SMRI at low shear values and MGSF at stronger shear values. The expressions of the MGSF criterion (Eqs.~\ref{MGSFcrit}) capture the limits of the unstable parameter domain. 

While the first corrective term, $\mathcal{C}_1$, shifts the minimum shear required for instability, the second term, $\mathcal{C}_2$, introduces additional stabilisation, particularly at low stratification and shear. This behaviour also holds in more viscous regimes, such as those corresponding to the DNS parameters listed in Table~\ref{table}. When the corrective terms $\mathcal{C}_1$ and $\mathcal{C}_2$ are omitted, the standard GSF criterion is recovered.

The criterion derived in this section can be used to assess the relevance of the MGSF instability in stellar interiors when the poloidal component of the magnetic field is known or can be reasonably estimated. A non-zero axial magnetic field ($\Lambda_Z \neq 0$) tends to stabilise configurations that would otherwise be unstable \citep[as noted by][]{Caleo2016GSF, Caleo2016Insta, Dymott2024}. 
\subsection{Magnetised GSF:\ Growth rate}
\label{subsec:mgsf_gr}

 \begin{figure}
\centering
\includegraphics[width=8.6cm]{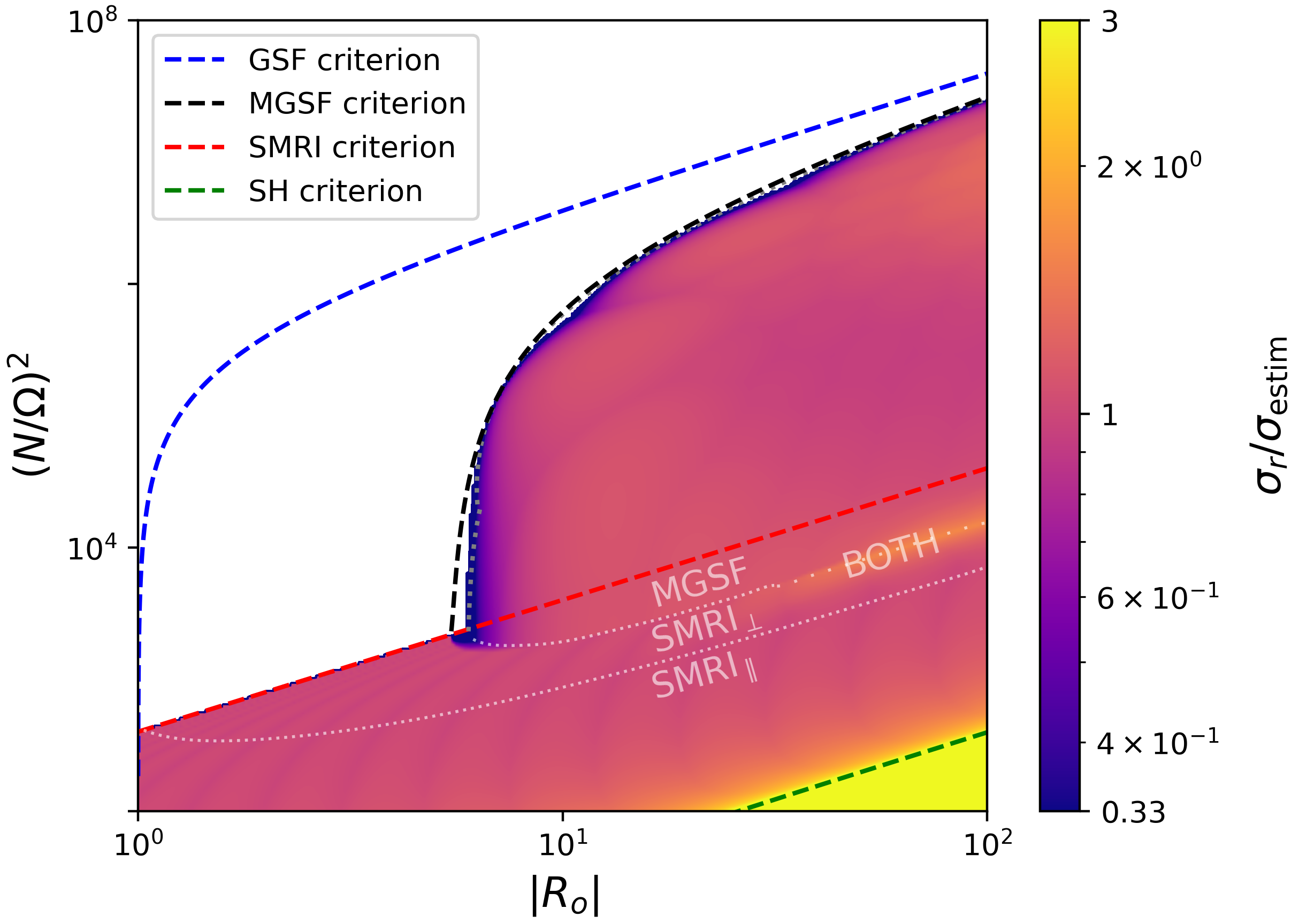} 
\caption{Ratio of the growth rate $\sigma_r$ to the analytical formula $\sigma_{\mathrm{estim}} \equiv \max(\sigma_{\mathrm{SMRI}}, \sigma_{\mathrm{MGSF}})$, which combines the predictions for SMRI (Eq.~\ref{gr_model}) and MGSF (Eq.~\ref{estimMGSF}). The flow parameters correspond to those of Fig.~\ref{sSMRI} (top panel), with $\Lambda_Z = 10^5$. The white dotted lines delimit the different instability domains (see Appendix~\ref{ApMGSF} and Fig.~\ref{kcri} for details). The black dashed line indicate the sufficient MGSF stability criterion (Eq.~\ref{MGSFcrit}). The grey dotted line correspond to the combination of the necessary criteria (Eq.~\ref{MGSFnece}). See Appendix~\ref{mgsf_criteria} for details.}
\label{GSFcri}
\end{figure}

To go further, we aim to derive a simple estimate of the growth rate of the magnetised GSF instability, enabling an assessment of its efficiency in radiative stellar regions. Considering regimes that are GSF-unstable but MRI-stable, we examine how the GSF growth rate varies with the flow parameters. As in the previous section, we find an approximate formula given by 

\begin{equation}
\begin{aligned}
\sigma_{\mathrm{MGSF}} \equiv \dfrac{\left(1-\mathcal{C}\right)^{3/4}}{1+2\sqrt{\vert \ross \vert} \mathcal{B}} \left(1 + 2 \dfrac{\Lambda_Z}{\sqrt{\vert \ross \vert -1}} \right)^{-1} \sigma_{0},
\end{aligned}
\label{estimMGSF}
\end{equation}
where 
\begin{equation}
\mathcal{B} = \frac{\pran \stra}{4\left(|\ross| - 1\right)}, \;
\mathcal{C} = \mathcal{B} +
\frac{\mathcal{C}_1 + \mathcal{C}_2}{4\left(|\ross| - 1\right)}.
\label{coeffestimMGSF}
\end{equation}
\noindent
We note that in Eq.~(\ref{estimMGSF}), the first prefactor is mainly relevant for regimes close to the marginal stability, whereas the second reflects the general magnetic slowing.

We illustrate the validity of this analytical formula in Figs.~\ref{GR} and~\ref{GSFcri}, where we introduce $\sigma_{\mathrm{estim}} \equiv \mathrm{max}(\sigma_{\mathrm{SMRI}},\sigma_{\mathrm{MGSF}})$. Across the unstable parameter domain, $\sigma_{\mathrm{estim}}$ remain very close to the numerical results $\sigma_r$, supporting the validity of our formula in strongly magnetised flows. In the middle of the transition region, however, $\sigma_{\mathrm{estim}}$ tends to underestimate $\sigma_r$ because it only accounts for the maximum of SMRI and MGSF, whereas both instabilities can act simultaneously. Moreover, near and below the SH criterion, the additional instability is not considered when it converges more closely to $\sigma_{\mathrm{0}}$. 

The analytical formula (Eq.\ref{estimMGSF}) is consistent with the simulations of \citet{Caleo2016Insta}, which show that the damping time of transient modes in a magnetised background is inversely proportional to the magnetic resistivity. The local analysis of \citet{Dymott2024} likewise highlights the role of Ohmic dissipation in weakening magnetic stabilisation. To illustrate the need for this criterion, we consider the case of the \emph{Kepler} 56 target: using the results of \citet{Fellay2021} together with a magnetic diffusivity $\eta = \eta_{\odot}$ (as in \citet{Caleo2016Insta}), we obtained a very low stabilizing threshold for the magnetic field, $B_{\mathrm{MGSF}} \sim 5\mathrm{G}$. This underscores the importance of accounting for magnetic stabilisation when modelling GSF modes in stars, particularly in stellar evolution models that include GSF-driven angular momentum transport.

\section{Global approach}
 \label{sec:global}

\subsection{Methodology}

We now consider a flow between two differentially rotating cylinders, at a radius of $R_{in}$, $R_{out} = 2 R_{in}$ having different angular velocities, $\Omega_{in}$, $\Omega_{out}$, and temperature, $T_{in}, T_{out}$,  immersed in a magnetic field, $\vec B = B_Z \vec e_Z + B_{\phi, in} (R_{in}/R) \vec e_\phi$. This problem is the magnetohydrodynamic, radially stratified version of the Taylor-Couette system. 
Such a configuration allows us to emphasise the effects of azimuthal curvature and radial boundary conditions and it is particularly suited to studying instabilities with short axial wavelengths (large $k_Z$), which are comparatively less sensitive to latitudinal curvature.

We extended the non-stratified linear stability code going back to \citet{Hollerbach2005, Hollerbach2010,Child2015}; we 
implemented a new temperature equation and radial temperature gradient as described in Appendix.~\ref{taylorcouette}. 
Then we chose to focus on modes that are confined within the bulk of the domain, excluding those that develop primarily near the inner or outer boundaries. To ensure this, we applied no-slip boundary conditions for the perturbed velocity field $\vec{U'}$, perfectly conducting boundaries for the magnetic field $\vec{B'}$, and fixed-temperature conditions for $T'$ (see Appendix~\ref{tc_profiles}). We assume perturbations to be of the form $A^\prime = a(R) \text{exp}[\sigma t + i(m\phi + k_Z Z)]$. We note that due to the large size of the matrix, $(5N_{rp}+12)^2$ with $N_{rp} \in [30,60]$, only viscous and weakly stratified regimes can be numerically solved. 

\subsection{Global MRI modes}

\subsubsection{MRI:\ Unstable parameter domain and growth rates}

\begin{figure}[ht]
\centering
\includegraphics[width=8.6cm]{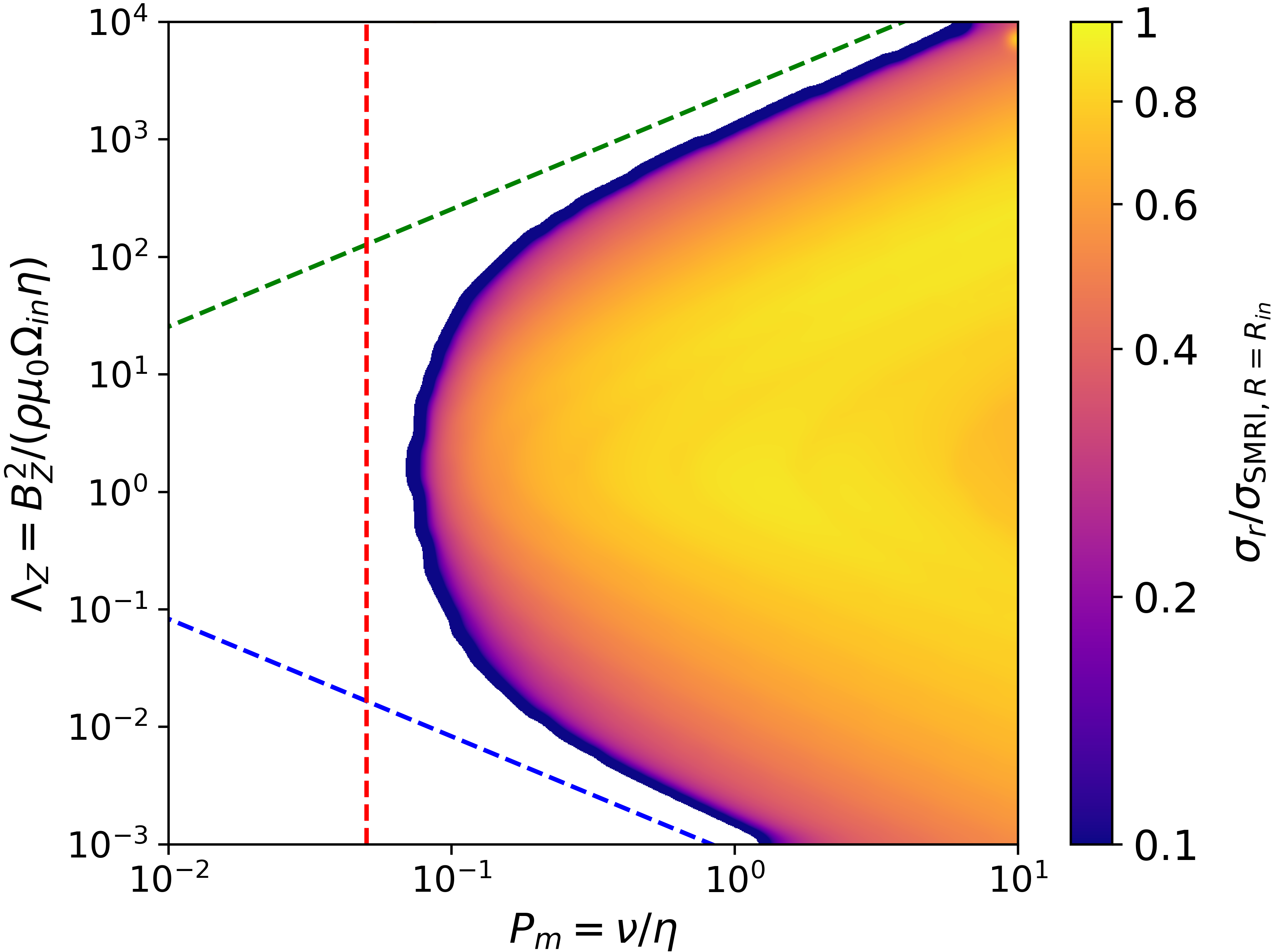}\\
\includegraphics[width=8.6cm]{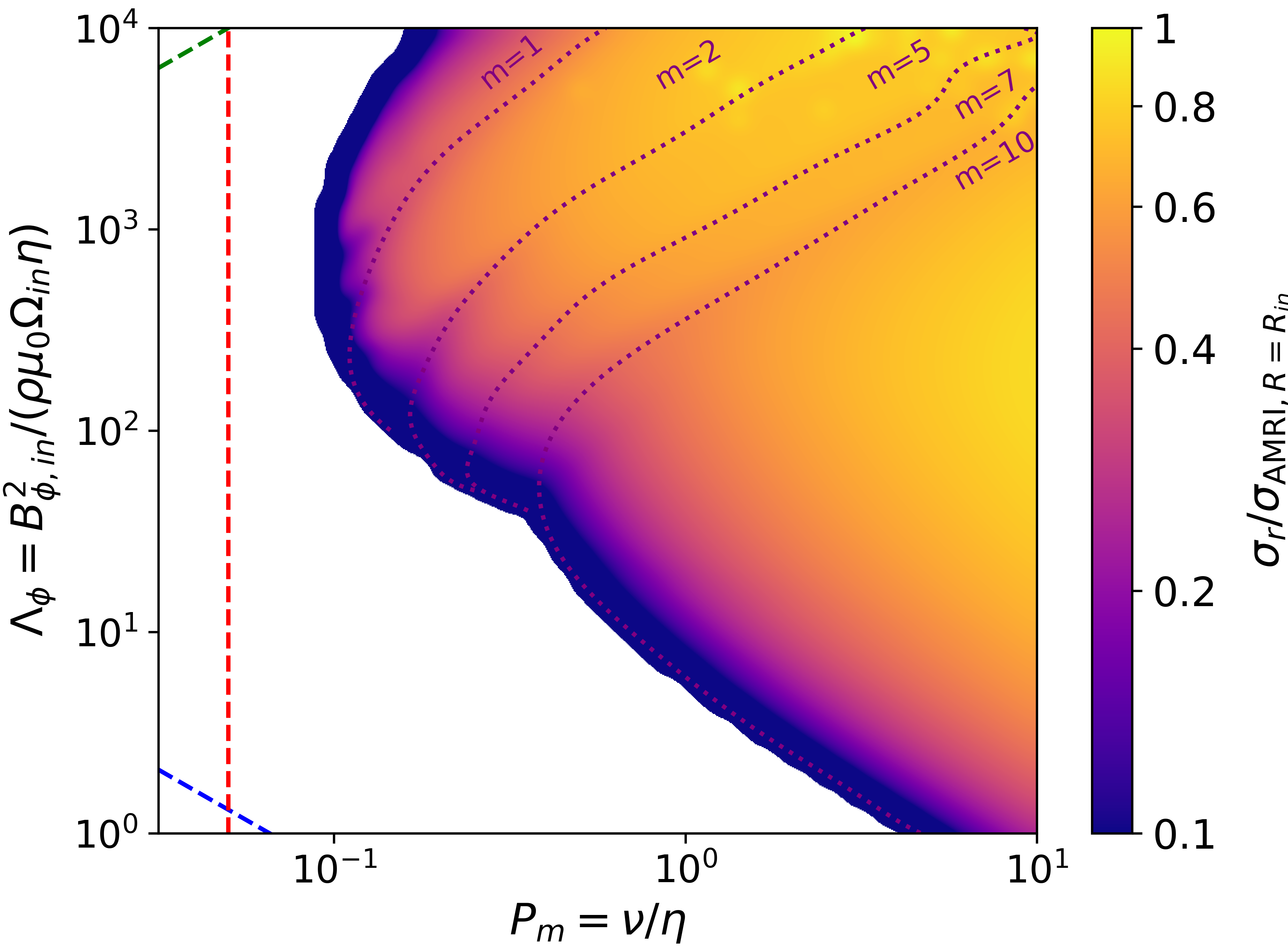}
  \caption{\textbf{Top panel:} Ratio $ \sigma_{\mathrm{r}}/\sigma_{\mathrm{SMRI}, R=R_{in}} $ in the $(\pram, \Lambda_Z)$ plane of the SMRI for regimes with $\pran=10^{-3}, \ekma = 10^{-5}, \ross=-0.5, \stra=10^2 $. The estimate $\sigma_{\mathrm{SMRI}, R=R_{in}}$ comes from the local model Eq.~\ref{gr_model} evaluated at $R=R_{in}$ and $\sigma_{\mathrm{r}}$ corresponds to the numerical values obtained with the global code. The green, red and blue dashed lines correspond to the criteria introduced in the local approach (Eq.~\ref{MRI_crit}). \textbf{Bottom panel:} Ratio $ \sigma_r / \sigma_{\mathrm{AMRI}, R=R_{in}}$ in the $(\pram, \Lambda_{\phi})$ plane of the Azimuthal MRI for the same regime than top panel and considering the fastest mode among the set of azimuthal wavenumbers $m\in [1,2,5,7,10]$ and $\sigma_{\mathrm{AMRI}, r=R_{in}}$ from Eq.~\ref{grAMRI2}. The blue and green dashed lines corresponds to the criteria for $m=1$ (Eq.~\ref{MRI_crit}). The dashed lines indicate the parameter region in which a given azimuthal mode number, $m$, yields the maximum growth rate.
 }
\label{MRI_glob}
\end{figure}

We first examined the conditions under which the SMRI arises in the $(\pram, \Lambda_{Z})$ plane, focussing on the flow regimes shown in Fig.~\ref{MRI_glob} (top panel). The unstable parameter domain closely matches the predictions of the local criterion (Eq.~\ref{MRI_crit}) and the growth rates, $\sigma_{\mathrm{r}}$, agree well with the analytical expression, $\sigma_{\mathrm{SMRI}}$ (Eq.~\ref{gr_model}), throughout the unstable region of Fig.~\ref{MRI_glob} (top panel). We also find a similarly good agreement for the axial wavenumbers, $k_Z$, of the modes, yielding remarkably consistent mode properties between the local and global analyses.

We then considered the AMRI in Fig.\ref{MRI_glob} (bottom panel) and examined the stability of regimes in the $(\pram, \Lambda_{\phi})$ plane for modes with $m \in [1, 2, 5, 7, 10]$. The expected scaling relations for thermal stratification and for the resistive limit were successfully recovered. However, the diffusive limit (Eq. \ref{MRI_crit}ii) departs significantly from the local prediction: in Fig.~\ref{MRI_glob} (bottom panel), $\Lambda_{\phi,\min}$ remains two orders of magnitude above the local estimate for $m=1$, and roughly four orders of magnitude above the local $m=10$ estimate (not shown). At low azimuthal wavenumber, $\Lambda_{\phi,\min}$ is almost independent of the diffusive ratio, $\pram = \nu / \eta$, in agreement with previous studies \citep{Rudiger2010, Rudiger2015}. We ruled out the variation of local background parameters as the origin of this discrepancy and find instead that the intrinsic non-axisymmetry of AMRI modes (e.g. curvature and azimuthal-drift effects) provides an additional stabilizing influence that is absent from the local WKB analysis \citep[for the low $\pram$ limit, see][]{Rudiger2014}. This could also explain why AMRI is more sensitive to global properties than SMRI. Whether this difference remains significant under stellar conditions is left for future investigation. However, within the unstable domain the growth rates $\sigma_r$ remain close to $\sigma_{\mathrm{AMRI}}$, except near the stability boundaries where additional non-local stabilizing effects become important. Overall, our results indicate that locally derived unstable domains and mode characteristics remain applicable to global SMRI and AMRI configurations, except in the regime of weak azimuthal fields, where additional stabilizing forces become significant.

\subsubsection{MRI:\ Mode location}

Beyond determining whether modes are unstable, it is crucial to predict where they preferentially develop within the flow. In stars, this identifies the regions that most efficiently transport angular momentum, mix elements, or amplify magnetic fields.

\begin{figure}[ht]
\centering
\includegraphics[width=8.6cm]{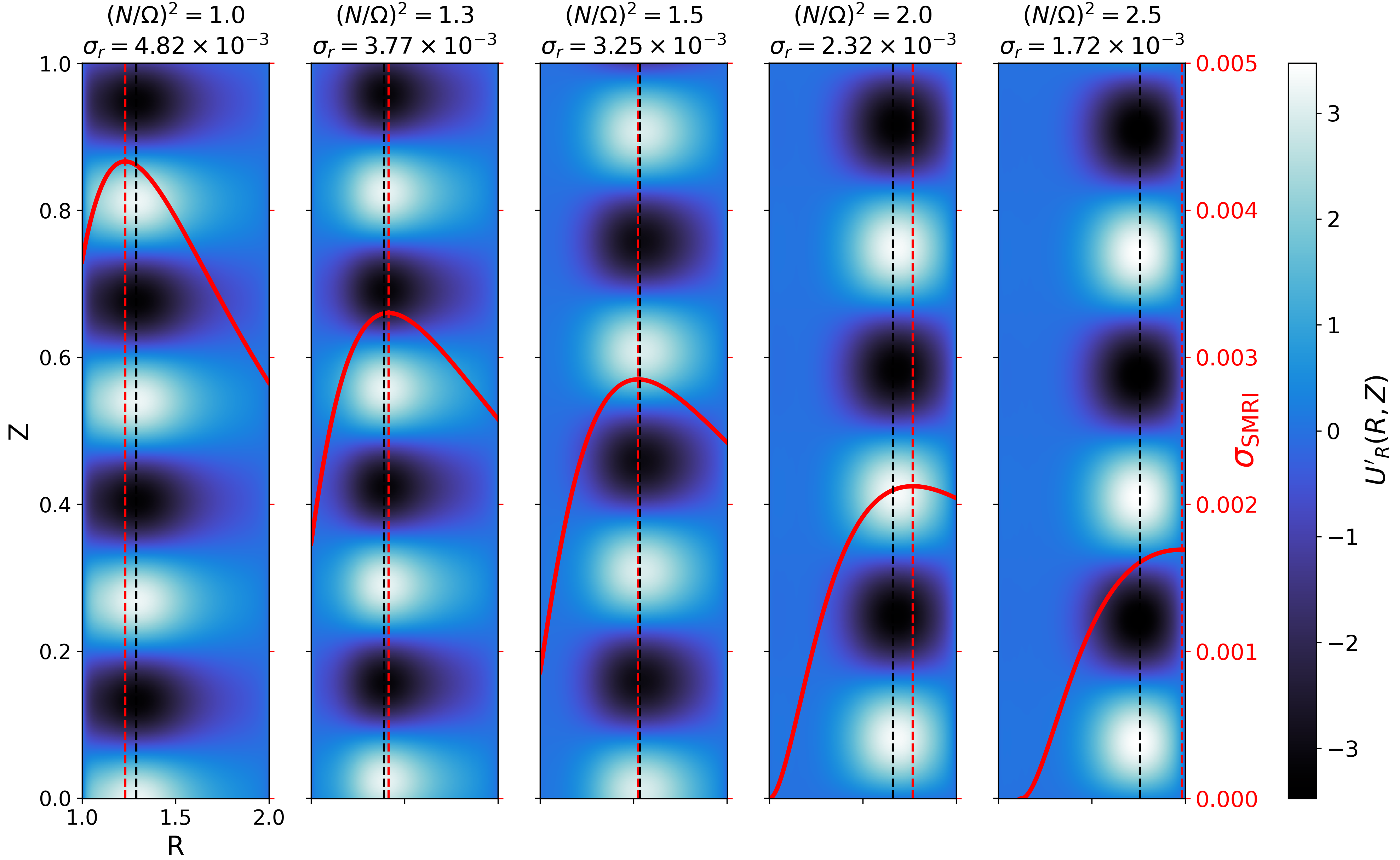}
  \caption{Reconstructed radial velocity perturbations $U'_R(R,Z)$ of the most unstable eigenmode in regimes with $\Lambda_Z = 1.6, \ekma=10^{-5}, \pram=1, \pran=0.1$, at $R=R_{in}$ and stratification amplitudes $\stra _{R=R_{in}}$ of $1, 1.3, 1.5, 2,$ and $2.5$. The maximal shear is $\vert \ross_{max} \vert= 0.05 $. The corresponding growth rate computed is given below the stratification value considered. The red overlayed curves correspond to the analytical formula (Eq.~\ref{gr_model}). The vertical black dashed lines give the radius where the mode amplitude peaks and the vertical red dashed lines give the radius where the analytical formula reaches its maximum.
          }
     \label{glob_SMRI}
\end{figure}

We first examined the location of SMRI modes between the two differentially rotating cylinders. Fixing the boundary temperatures and angular velocities through the diffusive equilibrium constrains the background profiles. Thus, we adopted the thermal stratification $\mathrm{N}^2(R) = \mathrm{N}^2_{R=R_{in}}(R_{in}/R)^4$ ad hoc to test cases with steep stratification. Figure~\ref{glob_SMRI} shows the radial velocity perturbations for different stratification strengths. The most unstable modes emerge at different radii, depending on the combined effect of shear and rotation, and they consistently localise where the analytical formula (red curves) peaks. The axial wavenumbers are slightly smaller than local estimates (see Table~\ref{tab:kglob_combined}). Finally, the radial extent of the modes is non-negligible, owing to diffusion, boundary effects, and the radial variation of the flow parameters, which sharpens the local growth rate profile.

Across different regimes and profiles studied, we find that the analytical formula of the SMRI growth rate (Eq.~\ref{gr_model}) also predicts the radial position of unstable SMRI modes. The numerical results for $\sigma_{\mathrm{r}}$ lie between $\sigma_{\mathrm{SMRI}}$ and $1.14 \times \sigma_{\mathrm{SMRI}}$, confirming the reliability of the local prediction. For the AMRI, the agreement remains satisfactory in several regimes, although curvature and boundary effects influence the instability location in some extreme parameter regimes. Additional configurations explored are presented in Appendix~\ref{loc}. While future works would need to validate this result with 3D global simulations, we conclude that taking the growth rates as estimated by local analysis constitutes a valuable proxy for estimating the location where the instability preferentially develops within the flow. 

\subsection{MGSF in a Taylor-Couette geometry}

Finally, we investigated the MGSF. The strong differential rotation required by this instability leads to large radial variations of $\Omega(R),$ which would induce spatial variations of the local dimensionless numbers. To retained a homogeneous force balance, we therefore prescribed constant values of $(\ross, \stra, \Lambda_Z, \ekma)$ throughout the domain, adjusting the coefficients so that the local parameters remain fixed. This choice isolates the role of magnetic stabilisation.

\begin{figure}[ht]
\centering
\includegraphics[width=8.6cm]{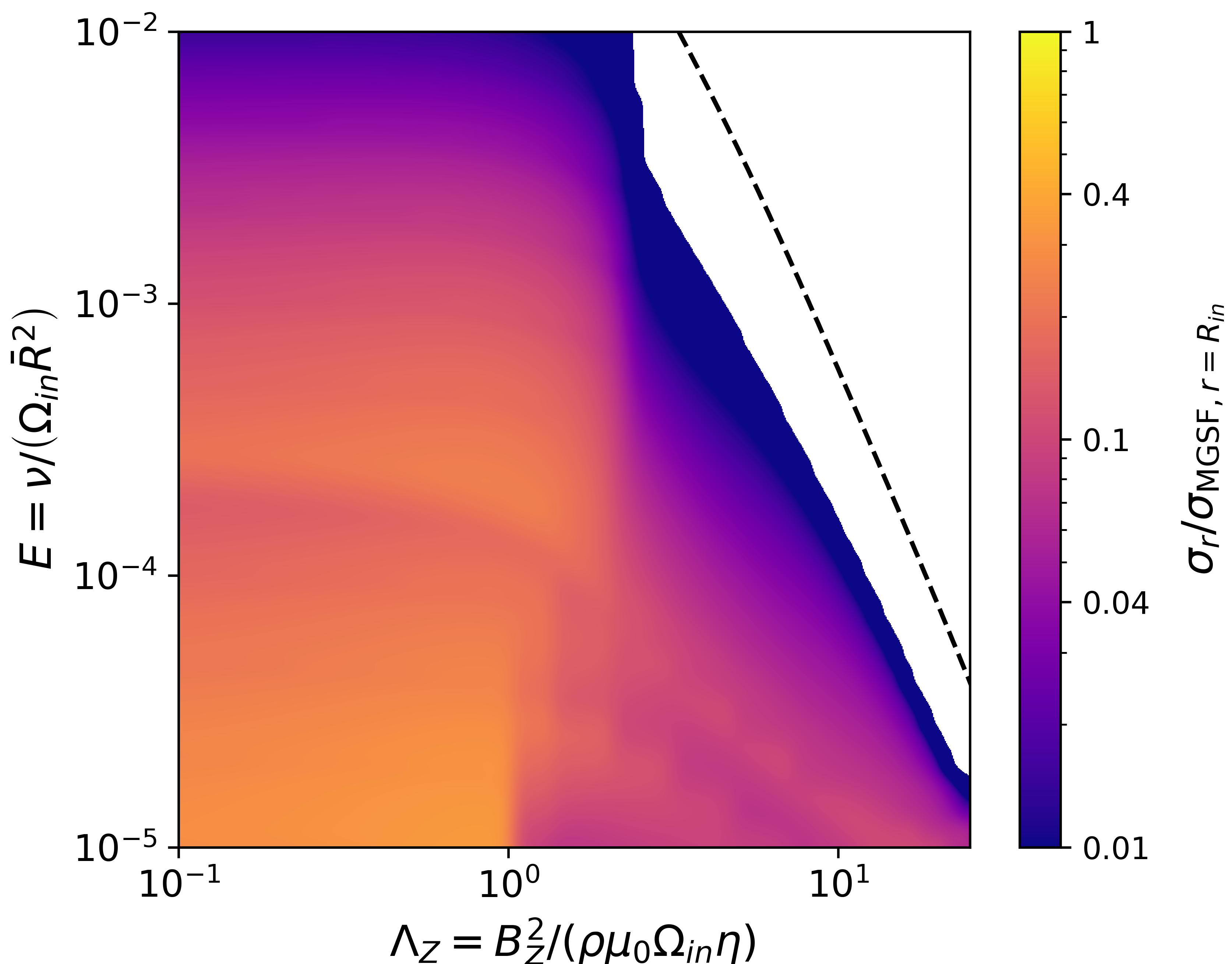}
  \caption{Ratio $\sigma_r / \sigma_{\mathrm{MGSF}, R=R_{in}}$ in the $(\Lambda_Z, \ekma)$ plane for the MGSF, with $\pram = 0.1$, $\pran = 10^{-2}$, $\ross=-6$, $\stra=400$, and $\Lambda_\phi = 0$ and the resolution, $N_{rp}=30$. Shear and stratification are constant across the domain, as are the Elsässer and Ekman numbers. The green dashed line marks the MGSF stability criterion (Eq.~\ref{MGSFcrit}).
  }
     \label{GSF_glob}
\end{figure}

Figure~\ref{GSF_glob} shows the phase diagram in the $(\Lambda, \ekma)$ plane for a flow configuration that is unstable to the GSF but stable to the MRI. Although numerical constraints limit our ability to explore stronger magnetic fields, the boundary of the unstable region agrees well with the extended criterion in Eq.~\ref{MGSFcrit}, which includes magnetic stabilisation. The eigenvalue problem is computationally demanding and sensitive to spurious modes: resolving GSF (and especially MGSF) modes requires fine radial discretisation\footnote{For the local approach to be valid for global modes, one requires $k_R R \gg 2 \vert \ross \vert$, where $k_R$ is the typical radial wavenumber of the modeand $\ross$ the local shear value. This condition translates into a numerical resolution requirement, $N_{rp} \gg 2 \vert \ross \vert$.} to obtain converged growth rates consistent with the local estimate (Eq.~\ref{estimMGSF}). However, increasing the resolution also reduces the numerical stability of the solver (see Appendix~\ref{sec:mgsf_numstab} for details). Despite these numerical challenges, the stabilising influence of magnetic fields is clearly reproduced: magnetic tension reduces the growth rate in accordance with the predicted scaling. Compared with the standard GSF, magnetised regimes involve smaller radial length scales, further reinforcing the need for high spatial resolution to accurately resolve MGSF modes.

The global Taylor-Couette analysis confirms the validity of the local approach. For the SMRI, both the stability thresholds and the growth rate estimates agree very well with the global results and they correctly predict the radial localisation of unstable modes across the domain. For the AMRI, some discrepancies appear in the weak-field regime, but the local estimates remain accurate throughout the unstable parameter domain. Finally, extending the analysis to strongly sheared flows shows that magnetic fields can effectively stabilise GSF modes.

\section{Application to low-mass subgiants and young red giants}
\label{sec:cesam}

We applied our analysis to stellar conditions representative of low-mass subgiants and young red giants. These evolutionary stages are particularly interesting because asteroseismic observations provide estimates of both core and envelope rotation rates \citep[hereafter D14]{Deheuvels2014}. Thus, our aim has been to determine whether fields comparable to those observed in red giant cores \citep[for a review, see][]{Deheuvels2024TASC} could give rise to shear-driven instabilities. We considered two stars from sample of D14 (denoted as A and F; see also Table \ref{table:sg_params}). 

\subsection{Stellar evolution models}

The stellar structure models used for this work were computed with the \cesamxx\footnote{\url{https://www.ias.u-psud.fr/cesam2k20/}} stellar evolution code (\citealt{Morel1997,Morel2008,Marques2013,Manchon2025}). The physical ingredients were chosen to be as close as possible to the one used by D14. Present models use \citet{Grevesse1993}'s determination of the solar chemical composition, with opacity tables from the OPAL team \citep{Rogers1992,Iglesias1996}, adapted to these abundances. The tables are supplemented, for low temperature values, by the Wichita\footnote{\url{https://www.wichita.edu/academics/fairmount_college_of_liberal_arts_and_sciences/physics/Research/opacity.php}} opacity tables \citep{Ferguson2005}. We note here that the chemical composition used by D14 to generate the opacity tables might be slightly different. The equation of state (\eos) uses data from the tables published as part of the OPAL 2005 \eos \citep{Rogers2002}. The nuclear reaction rates follow the compilations from NACRE \citep{Aikawa2006}, apart from LUNA \citep{Broggini2018} for the ${}^{14}{\rm N(p},\gamma){}^{15}{\rm O}$ reaction. The convection is modelled with the full turbulent spectrum formalism of \citet{Canuto1996}. The atmosphere was retrieved using the Eddington $T(\tau)$ relation. The effects of overshoot and rotation were not considered.

To recover observational parameters within uncertainties given by D14, we had to slightly adjust the input parameter of each model, within the range of values they considered for their grid-based approach. This is due to the different version of the code, small differences in the input physics, and round-off errors in the value of the input parameters provided in D14. These parameters are gathered in Table \ref{table:sg_params}. From the models, the profiles of Prandtl, $\pran = \nu / \kappa$, and magnetic Prandtl, $\pram = \nu/ \eta$, numbers have been extracted, where $\nu$ is the kinematic viscosity, $\eta$ the molecular magnetic diffusivity, and $\kappa$ the thermal diffusivity, following the formulations of \cite{Spitzer1962,Mihalas1984}.

\begin{table*}[t]
\begin{center}
\caption{Input parameters and surface properties of the stellar models of stars A and F in the D14 stellar sample,  computed with the newest version of \cesamxx.}
\begin{tabular}{l l l l l l l l}
\hline\hline                                                                         \\[-7pt]
Star             & $M$    & Age    & $(Z/X)_0$ & $Y_0$ & $\alpha_{\rm CGM}$ & $T_{\rm eff}$ & $\log g$ \\
                 & $[M_\odot]$     & [Gyrs]    & --    & --    & --         & [K]           & -- \\ \hline
A (KIC 12508433) & 1.216  & 5.86   & 0.050     & 0.3   & 0.6   & 5292       & 3.85 \\
F (KIC 9574283)  & 1.085  & 6.04   & 0.012     & 0.27  & 0.625 & 5117       & 3.58  \\
\hline
\end{tabular}
\label{table:sg_params}
\end{center}
\end{table*}

\subsection{Rotation and magnetic field models}

Asteroseimic observations provide estimates of mean core and surface rotation so that the exact profile is not accurately know. Therefore, based on the core and envelope rotation-rate estimates given in D14, we constructed two types of models, which we then tested. A smooth one as proposed by D14,
\begin{equation}
\Omega_{\mathrm{lin}}(r) = \Omega_{in} + (\Omega_{out}-\Omega_{in})\dfrac{r}{r_{\star}},
\end{equation}
\noindent
 with $r_\star$ the radius of the radiative-convective transition and a sharper one assuming a strong decrease of angular velocity \citep[see also][]{Belkacem2015} set with
\begin{equation}
\Omega_{\mathrm{sharp}}(m) = \Omega_{in} + \dfrac{\Omega_{out}-\Omega_{in}}{2}\left(1+\text{erf}\left(\dfrac{m-m_d}{w_d}\right)\right),
\end{equation}
\noindent
where $m$ is the local mass, $m_d$ defines the mass where the angular velocity drops, and $w_d$ controls the width of the transition.

For the magnetic field, several estimates of the core magnetic field of red giants have been debated \citep[e.g.][]{Fuller2015, Mosser2017, Li2022, Deheuvels2023}. For some of them, strong radial fields of the order of $100\,\mathrm{kG}$ have been detected near the hydrogen-burning shell. In the following, we assume a poloidal magnetic field dominated by a dipolar component, such that the axial and the radial field components are of the same order, $B_Z \sim B_r$. We consider the axial component at the equator of the magnetic field generated by a dipole, expressed as 
\begin{equation}
B_{Z}(r) = B_{Z,0} \left(\dfrac{r_b}{r}\right)^3,
\end{equation}
\noindent
where $B_{Z,0}$ is the axial field value at the radius, $r_b$, where the Brunt-Väissälä frequency reaches its maximum. This approach allows us to use instabilities onset properties expressed in terms of $B_{Z,0}$, while discussing the astrophysical case, where only the radial field $B_r$ is observationally constrained near the hydrogen-burning shell. The case of an azimuthal or toroidal magnetic field will be addressed in a forthcoming study, where (in addition to the AMRI) magnetic kink-type instabilities such as the Tayler instability (TI) are expected to play a role.

Then, we can solve Eq.~\ref{disp}, using the local background values at each radius to determine the instability domains predicted by our local approach. 
For all models tested, we found that the SMRI and MGSF criteria (Eqs.~\ref{MRI_crit} and \ref{MGSFcrit}), together with the corresponding analytical expressions for their growth rates (Eqs.~\ref{gr_model} and \ref{estimMGSF}), successfully identify the unstable regions and accurately characterise their properties. We emphasise that these semi-analytical formulas (Eq.~\ref{tau_model}) reliably predict both the existence and the timescales of the instabilities at very low computational cost (without the need to solve Eq.~\ref{disp}) and can therefore be implemented directly in stellar evolution codes. When the regime is unstable, the corresponding growing times are expressed as

\begin{equation}
\begin{aligned}
\tau_{\mathrm{SMRI}} &\equiv \tau_0 \left[1+Q^2\left(\dfrac{\Lambda_Z}{\Lambda_{Z,U}} + \dfrac{\Lambda_{Z,L}}{\Lambda_Z}\right) \right],  \\
 \tau_{\mathrm{MGSF}} &\equiv  \tau_0 \left(1+ \dfrac{2 \Lambda_Z}{\sqrt{\vert \ross \vert -1}}\right) \left( \dfrac{1+2\sqrt{\vert \ross \vert} \mathcal{B}}{\left(1-\mathcal{C}\right)^{3/4}} \right), 
\label{tau_model}
\end{aligned}
\end{equation}
\noindent

with $\tau_0, Q, \Lambda_{Z,L}, \Lambda_{Z,U},  \mathcal{B}, \mathcal{C}$ as presented in Sects.~\ref{sec:methodo} and~\ref{sec:GSF}. \color{black}

\subsection{Results} 

For evolved low-mass stars, structural evolution is essentially driven by the core contraction with a typical timescale, denoted as $\bar \tau_{\rm exp}$. 
The spatial mean of this characteristic timescale is found for stars A and F to be,   \(\bar{\tau}_{\rm exp, A} \sim 425\ \mathrm{Myr}\) and \(\bar{\tau}_{\rm exp, F} \sim 239\ \mathrm{Myr, }\) respectively. Comparing the instability growth times with this characteristic timescale provides a measure of whether (and to what extent) these instabilities can develop during the stellar evolution phase. 

\begin{figure*}[h]
\centering
\includegraphics[width=8.5cm]{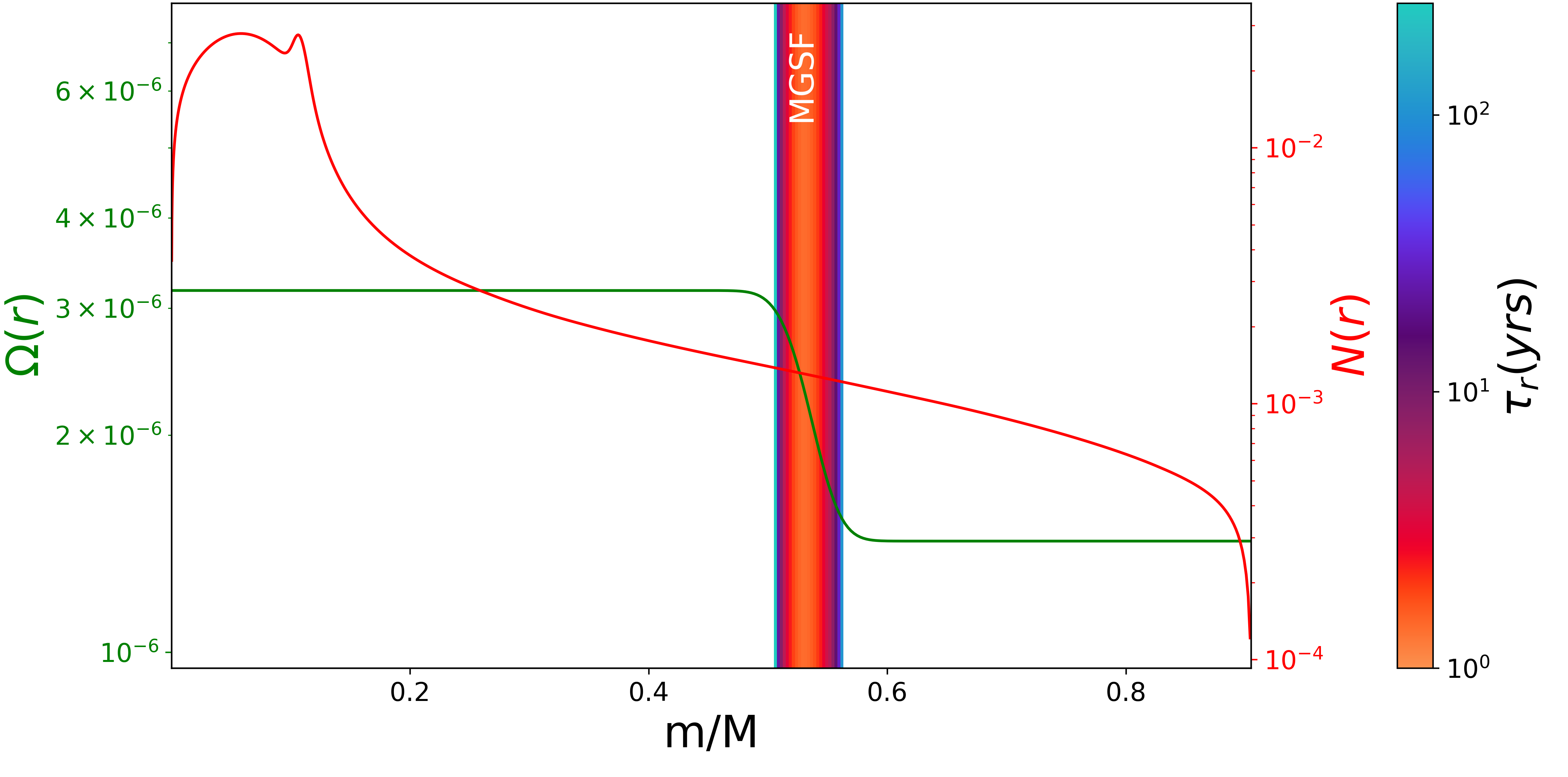}
\includegraphics[width=8.5cm]{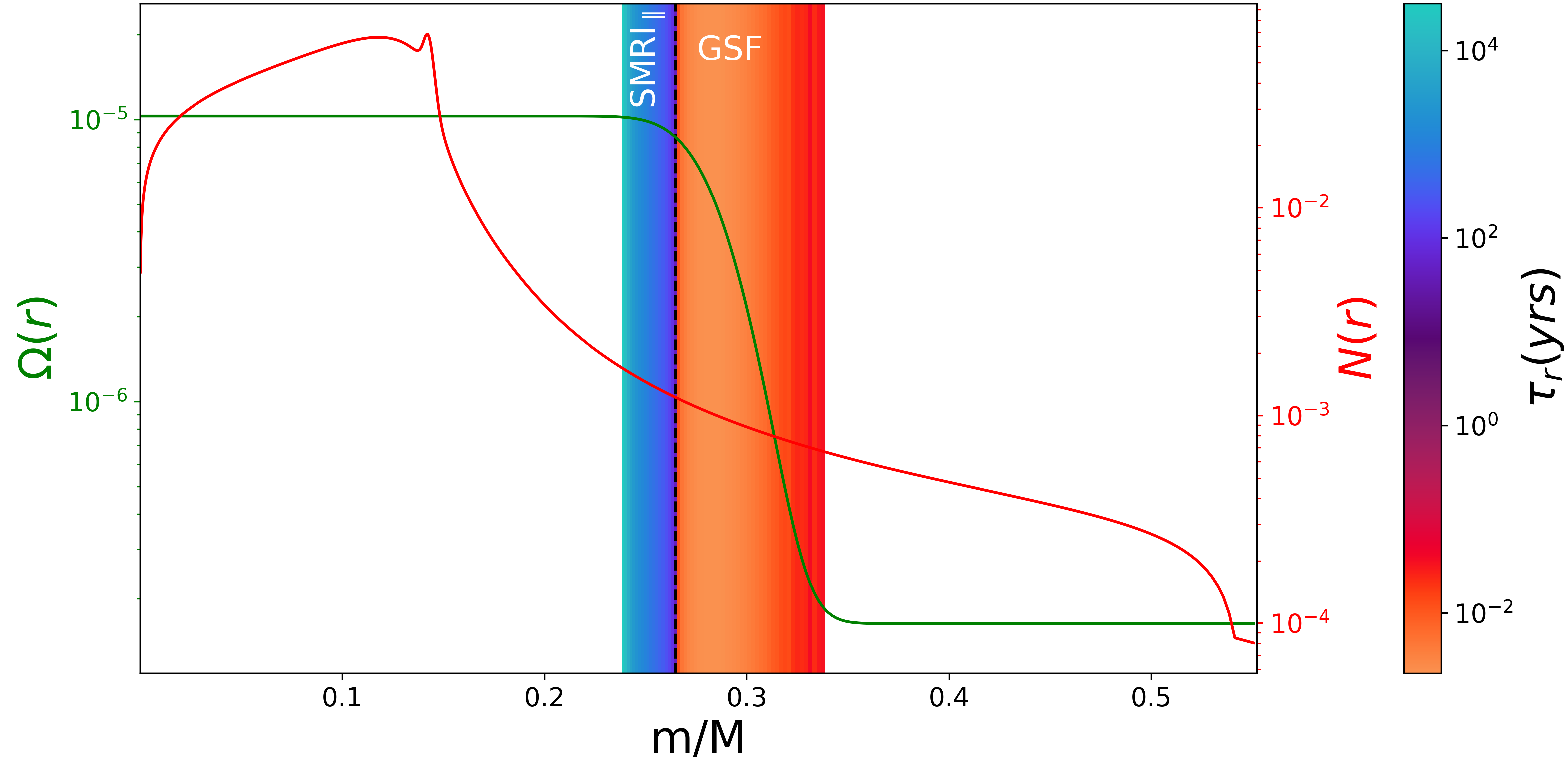}
\includegraphics[width=8.5cm]{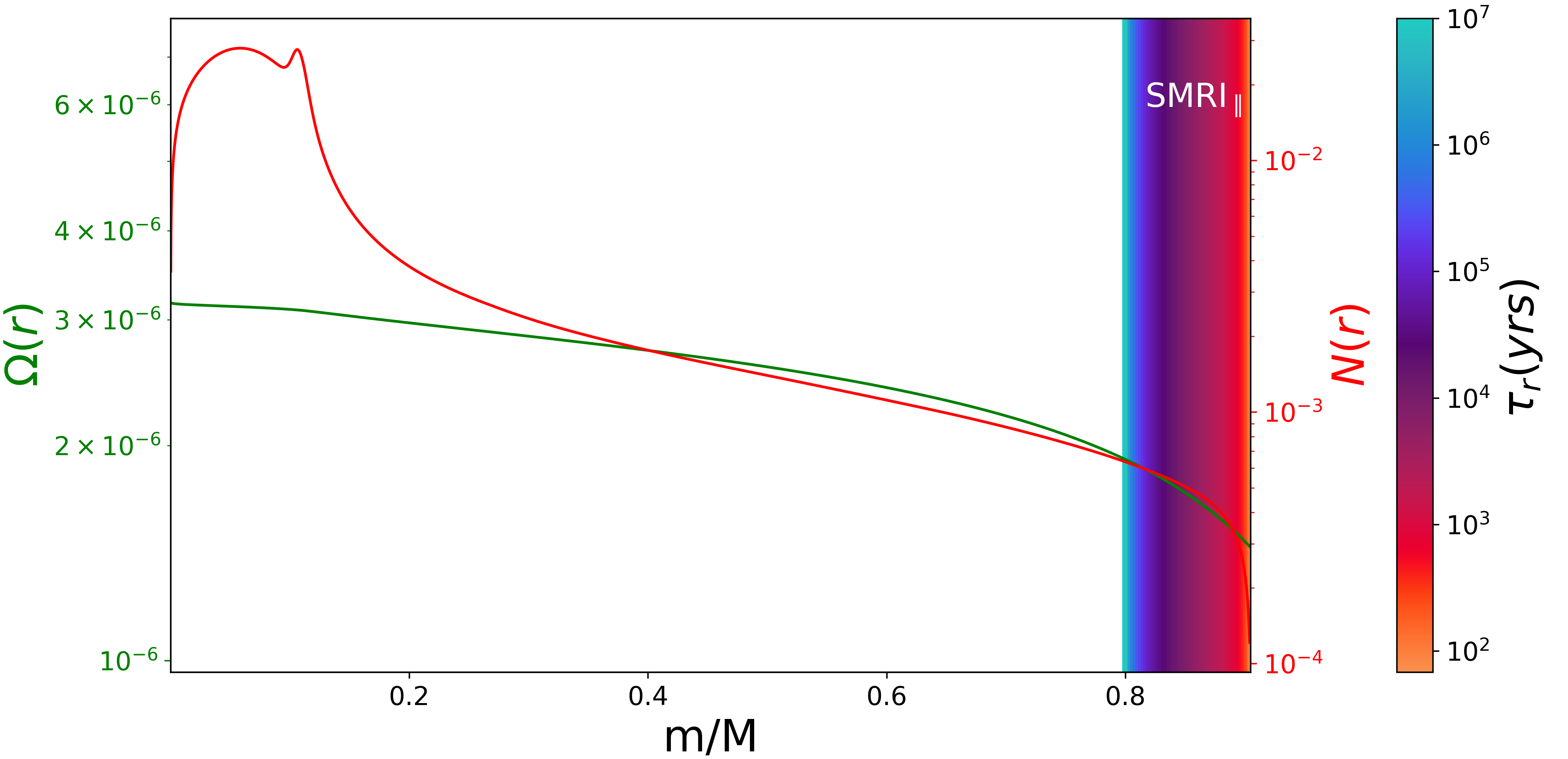}
\includegraphics[width=8.5cm]{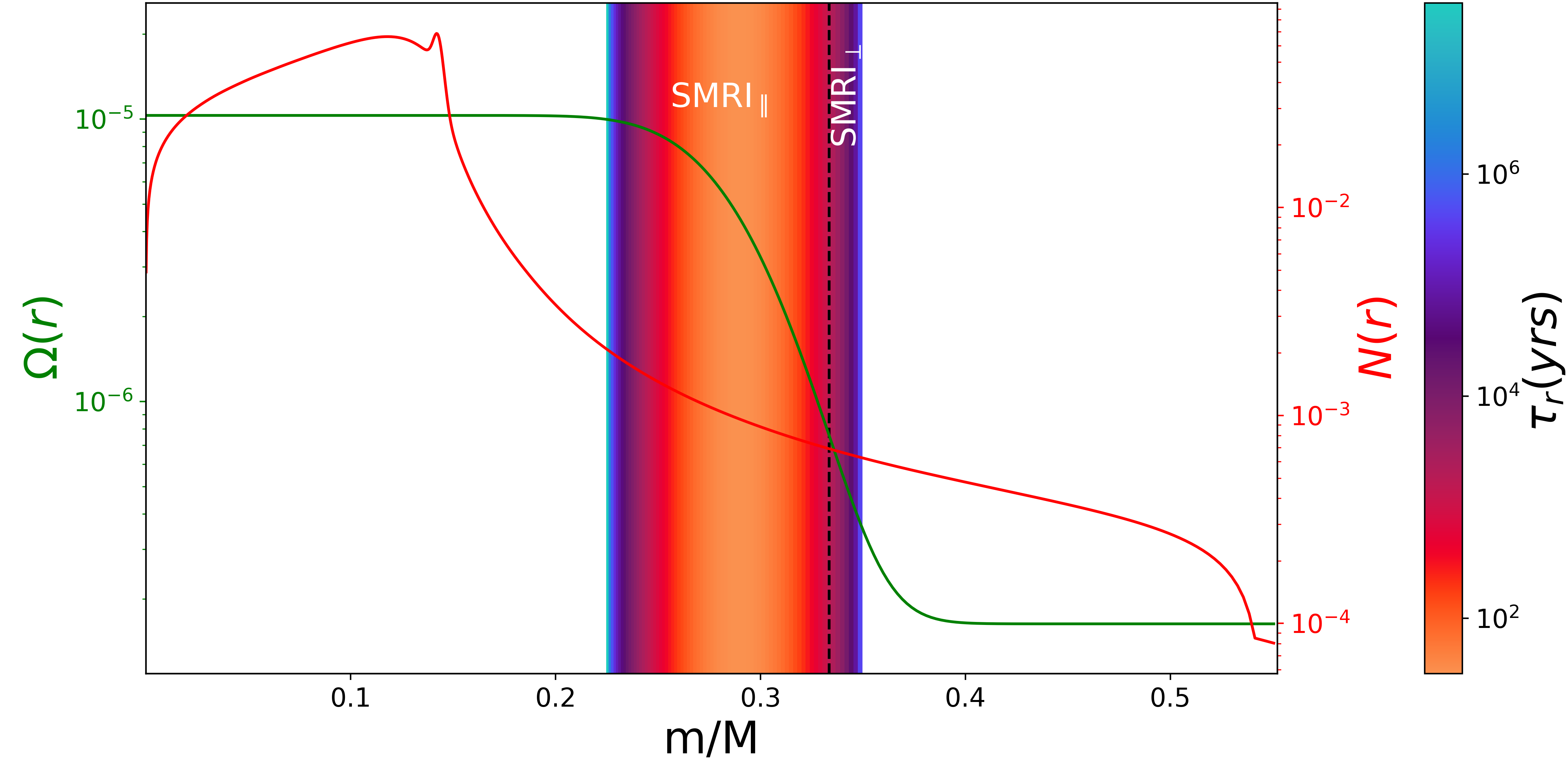}
\caption{Growth rates of unstable regions for different rotation and Brunt-Väisälä (BV) frequency profiles, shown in the background color scale, are displayed for the stars considered. The type of shear-driven instability (e.g. SMRI, MGSF) is indicated in each panel. The BV profile exhibits a narrow peak corresponding to the hydrogen-burning shell. Its position defines the reference radius, $r_b$. The reference width, $w_b$, is defined as the radial distance over which the BV frequency decreases by an order of magnitude from its maximum at $r_b$.
  \textbf{Left panels:}  Star A,  \textit{top:} $B_{Z,0} = 1\,\mathrm{kG}, w_d = w_b, r_d = 5 r_b$. \textit{Bottom:}  $B_{Z,0} = 100\,\mathrm{kG},$ linear rotation profile $\Omega(r)$.
  \textbf{Right panels:} Star F, \textit{top:} $B_{Z,0} = 1\,\mathrm{G}, w_d = 3 w_b, r_d = 2 r_b$. \textit{Bottom:}  $B_{Z,0} = 100\,\mathrm{kG}, w_d = 5 w_b, r_d = 2 r_b$.
          }
\label{Ces}
\end{figure*}

\vspace{1mm}
\textbf{Star A:}
In Fig.~\ref{Ces} (left panels), we present the results for Star~A. The top panel corresponds to a sharp rotation profile and a moderate magnetic field ($B_{Z,r=r_b}=1\,\mathrm{kG}$), showing the development of the MGSF. For stronger fields, magnetic stabilisation progressively suppresses the MGSF, leaving the SMRI as the dominant instability, while for weaker fields the GSF prevails \footnote{In the weak magnetic field limit, the MGSF corresponds to the standard GSF, the stability criteria, growth rates, and mode wavenumbers are poorly affected by the magnetic field.}.
When a strong magnetic field is considered ($B_{Z,r=r_b}=100\,\mathrm{kG}$), the growth time is $\tau_r\sim 46\,\mathrm{kyr} \ll \bar \tau_{\mathrm{exp,A}}$, suggesting that the SMRI could develop within the thin shear region.
For the model with a linear rotation profile (bottom panel), only the SMRI remains unstable and only in the outer layers where stratification is relatively weak. In this case, the shear lies in a weakly magnetised region, even when assuming a strong field of $B_{Z,r=r_b}=100\,\mathrm{kG}$ in the hydrogen-burning shell. Consequently, the growth time decreases when going from $B_{Z,r=r_b}=1\,\mathrm{G}$ to $100\,\mathrm{kG}$, reaching $\tau_r\sim 149\,\mathrm{yr}$ in the upper radiative layers and rising sharply with depth.
From the other models tested and summarised in Table~\ref{table:resucesam} (Appendix~\ref{cesamodels}), we conclude that for the subgiant Star~A, the development of a shear-driven instability requires the shear region to be both narrow enough to sustain a strong rotational gradient and located sufficiently far from the hydrogen-burning shell to mitigate thermal and magnetic stabilisation.

\vspace{1mm}
\textbf{Star F:}
In Fig.~\ref{Ces} (right panels), we present the result for Star~F. This more evolved star exhibits a larger rotation contrast, $\Omega_{in}/\Omega_{out}$, implying stronger shear and a flow more prone to shear-driven instabilities. In the top panel, we consider a sharp rotation profile and a weak magnetic field ($B_{Z,r=r_b}=1\,\mathrm{G}$). In the bottom panel, we adopted a smoother profile with a stronger field ($B_{Z,r=r_b}=100\,\mathrm{kG}$).
For the sharp profile, both SMRI and GSF can develop in distinct regions, with the GSF growing on the fastest  timescale, $\tau_r \sim 20 \, \rm hr$. When the magnetic field exceeds $1\,\mathrm{kG}$, however, only the SMRI persists, and its growth rate decreases significantly. With the smoother rotation profile and $B_{Z,r=r_b}=100\,\mathrm{kG}$, our analysis predicts SMRI growth on a timescale of $\tau_r \sim 31.8 \,\mathrm{yr}$. Shear regions located closer to the hydrogen-burning shell experience stronger stabilisation from stratification and magnetic tension, reducing their susceptibility to instability. Conversely, shear zones further from this shell are less stabilised, favouring the growth of shear-driven modes. This again explains why smoother (linear) rotation profiles are particularly prone to fast-growing instabilities in their outer radiative regions.

Overall, if a strong large-scale poloidal magnetic field of the order $100\,\mathrm{kG}$ is present in the hydrogen-burning shell of a low-mass subgiant or young red giant (as observed in some red giants) shear-driven instabilities can develop efficiently only if the shear region lies sufficiently far from that shell. Interestingly, for all tested models, the instabilities found have a growth time much lower than the expansion timescale. Additionally, we found only SMRI modes and no GSF or MGSF, when $B_{Z,r=r_b} \geq 1\,\mathrm{kG}$.

Although the degree of differential rotation can vary during the post-main sequence evolution, it generally becomes stronger in the early red giant phase for the models considered here. Consequently, shear-driven instabilities are expected to grow more rapidly in young red giants than in sub-giants. For the different rotation and magnetic field configurations tested, we found growth times ranging from a few months to several hundred thousand years, highlighting the need to account for magnetic and stratification effects when assessing the role of shear-driven instabilities in stellar evolution. The simple estimates proposed here can be implemented in 1D stellar evolution codes to evaluate, for instance, the efficiency of angular momentum transport by SMRI through an effective viscosity, $\nu_{\mathrm{SMRI}} = \sigma_{\mathrm{SMRI}} R^2$, where the instability develops at radius, $R$, with a growth rate of $\sigma_{\mathrm{SMRI}}$ \citep[similarly to][]{Wheeler2015, Griffiths2022}. Since GSF and MGSF modes are found to be nearly perpendicular, transport prescriptions derived from previous studies of the GSF instability \citep[e.g.][]{Barker2019} might not be directly applicable in the magnetic case. While our linear results may provide qualitative guidance on the influence of magnetic fields, dedicated non-linear simulations with a sufficiently fine mesh would be required to establish a saturation theory for the MGSF and, in turn, a reliable model for the associated angular momentum transport.

\section{Conclusion}
\label{sec:conclusion}
We investigated shear-driven instabilities in stellar radiative regions through a linear stability analysis of MRI and GSF modes. Using a WKB approach, we first derived general estimates for the local unstable modes, then extended the analysis to global Taylor-Couette configurations to assess how these modes arise in finite domains. Finally, we applied our results to stellar models of subgiants and young red giants. 

For the SMRI and AMRI, we recovered already known stability criteria and quantified the influence of stratification, magnetic tension, and diffusion on growth rates. In strongly sheared regimes, we derived a new criterion for the magnetised GSF (MGSF) instability and estimated how magnetic and stratification effects reduce the range of unstable modes, clarifying the transition between SMRI and MGSF. 

In addition, we derived approximate formulae for the instability growth times, making it possible to determine which instability (i.e. SMRI or MGSF) is likely to dominate under given stellar conditions. Implementing these formulae in 1D stellar evolution codes, through the effective viscosity associated with these instabilities, would enable a more realistic estimation of angular momentum transport throughout stellar evolution. Our results were further validated through stability analyses of Taylor-Couette flows. The global modes confirm that the local WKB framework accurately identifies where instabilities develop and provides reliable estimates of their behaviour across the domain. This strengthens the applicability of the local results to realistic stellar profiles.  

When applied to sub-giants and young red giants, our analysis shows that shear-driven instabilities can grow on short timescales for magnetic fields well below $100\,\mathrm{kG}$, suggesting that these mechanisms may readily operate in such stars. Depending on the magnitude and radial location of the shear, our analytical formulae predict where SMRI or MGSF modes should arise within the radiative zone. In contrast, strong axial magnetic fields of order $\sim 100\,\mathrm{kG}$ confined to the hydrogen-burning shell can co-exist with shear-driven instabilities only if the shear layer lies sufficiently far from the shell; in such cases, magnetic tension and stratification act together to suppress most unstable modes. Overall, these results motivate incorporating our instability criteria and growth rate estimates into 1D stellar evolution models to evaluate the role and efficiency of shear-driven instabilities throughout stellar evolution.

A follow-up paper will address magnetically driven instabilities, focussing on Tayler-like modes \citep{Tayler1973, Spruit1999, Skoutnev2024a}, thereby complementing recent work on Tayler dynamos \citep[e.g.][]{Petitdemange2023, Meduri2025} demonstrating efficient angular momentum transport. The interplay between the MRI, MGSF and Tayler instabilities and, more generally, the potential co-existence of multiple instabilities across different spatial or temporal scales remain an open question \citep{Jouve2020, Chang2021, Meduri2024a}. In addition, the present approach primarily describes the linear onset of the instabilities and cannot independently constrain their non-linear saturation or the associated angular momentum transport, which will be further addressed in future non-linear simulations.

--------------------------------------------
\begin{acknowledgements}
The authors thank Oleg Kirillov and Jordan Philidet for useful discussions that helped clarify the work presented, and the anonymous referee for their relevant remarks. They also acknowledge financial support from the "Initiative Physique des Infinis" (IPI), Sorbonne Université, and the "Action Thématique de Physique Stellaire" (ATPS).
\end{acknowledgements}

\bibliographystyle{aa}
\bibliography{bibliography/biblio_virgin}

\begin{appendix}
\section{Method for local linear analysis}
\label{num_met}

We solve the eigenvalue problem (Eq.~\ref{disp}) for various set of flow parameters $(\ekma, \Lambda, \Lambda_\phi, \pran, \pram, \stra, \ross)$ (see Table~\ref{table}) and different modes ($k_R, k_Z$) up to $100 \times 100$ 
values in $([2 \pi; 10^8], [2 \pi; 10^8])$ when necessary. For a given parameter regime, we seek to find the optimal mode, for a given azimuthal number $m$, which maximises the value of the growth rate because we expect the instability to be driven by the most unstable one. The eigenproblem is solved using the Durand-Kerner method \citep{Durand,Kerner}, which is a simultaneous multi-fixed-points iteration algorithm. The principle is as follows: consider $P_{\mathcal{H}}$ the $5^{th}$ order characteristic polynomial associated to our problem, its roots $\sigma_{i}$ and $\sigma_{0,i}$ the initial guesses of those roots. Concerning the initial guesses, we start with a set of arbitrary values, for subsequent calculations the initial guesses $\sigma_{0,i}$ can be selected from previously calculated roots corresponding to similar flow regime parameters and mode values. Then, the algorithm works by iteration for each $i$, the guesses at the step $k$ are obtained with the relation: $\sigma_{k,i} = \sigma_{k-1,i} - P_{\mathcal{H}}(\sigma_{k-1,i})/\prod_{j , j \neq i} (\sigma_{k-1,i} - \sigma_{k-1,j})$. 

Then, to study the role of two parameters on a domain of instability, we construct a mesh for those two parameters of interest, using a $300 \times 300$ grid of logarithmically spaced values. For each point in this grid, corresponding to a specific set of flow regime parameters, we solve the eigenvalue problem (Eq. \ref{disp}).

\section{MRI eigenmode properties}
\label{ApMRI}

\begin{figure}[ht]
\centering
\includegraphics[width=9cm]{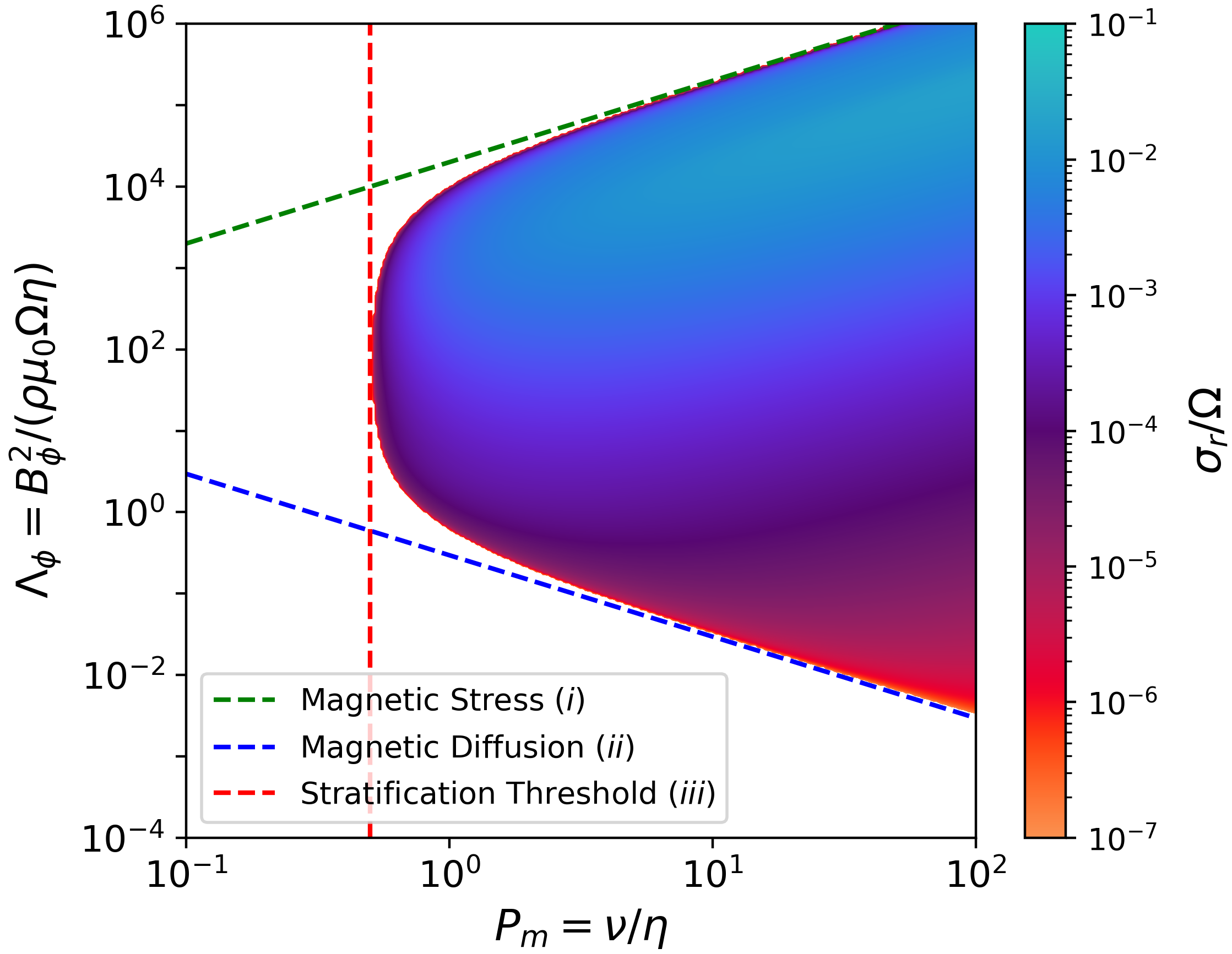}
\includegraphics[width=9cm]{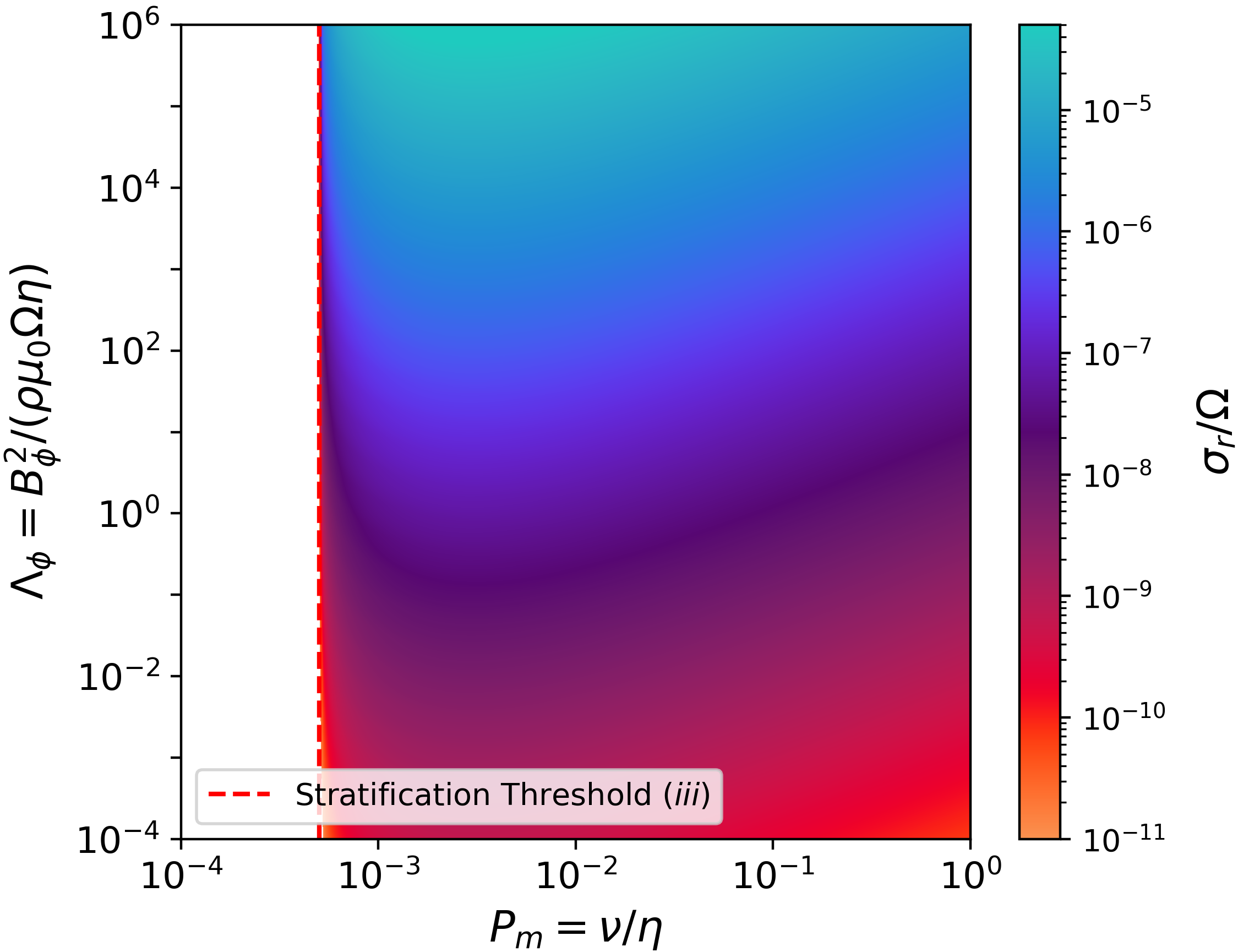}
  \caption{ \textbf{Top panel} Growth rate $\sigma_r$ normalised by $\Omega$ in the $(\pram, \Lambda_\phi)$ plane of the axisymmetric AMRI, for a regime typical of DNS (see Table~\ref{table}, with  $\vert \ross \vert = 0.05, \Lambda_Z = 0, \stra = 1, m=1$). The three criteria (Eq.~\ref{MRI_crit}) correspond to the limits of the unstable parameter domain. The blank region corresponds to the stable parameter domain. 
  \textbf{Bottom panel} Same study but for the stellar radiative regime set of parameters, $\vert \ross \vert = 0.05, m=1$ and $\stra = 10$.
  }
     \label{amri}
\end{figure}

In Fig.~\ref{amri}, we display AMRI unstable parameter domains and their corresponding growth rate values, for both regimes listed in Table~\ref{table}, in the phase diagram  $(\pram, \Lambda_\phi)$ with $\vert \ross \vert = 0.05$ and $\stra = 1$ and $10$, respectively for the DNS regime and radiative stellar regime. We find perfect agreement with the criterion given in Eq.~\ref{MRI_crit}, as in the case of the SMRI, and the AMRI growth rate estimate from Eq.~\ref{grAMRI2} shows good agreement with the numerical values.

\begin{figure}[ht]
\centering
\includegraphics[width=9cm]{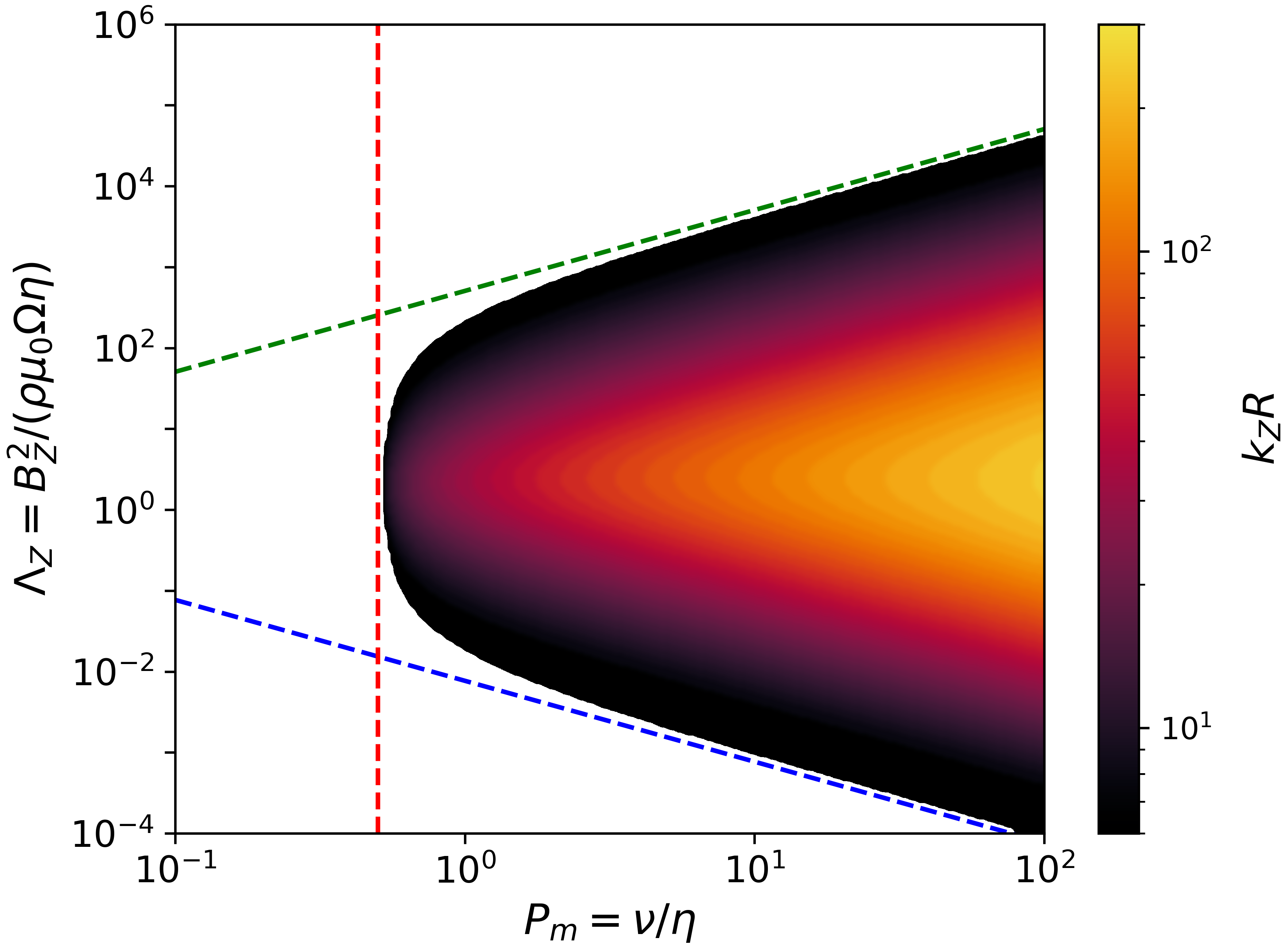}
\includegraphics[width=9cm]{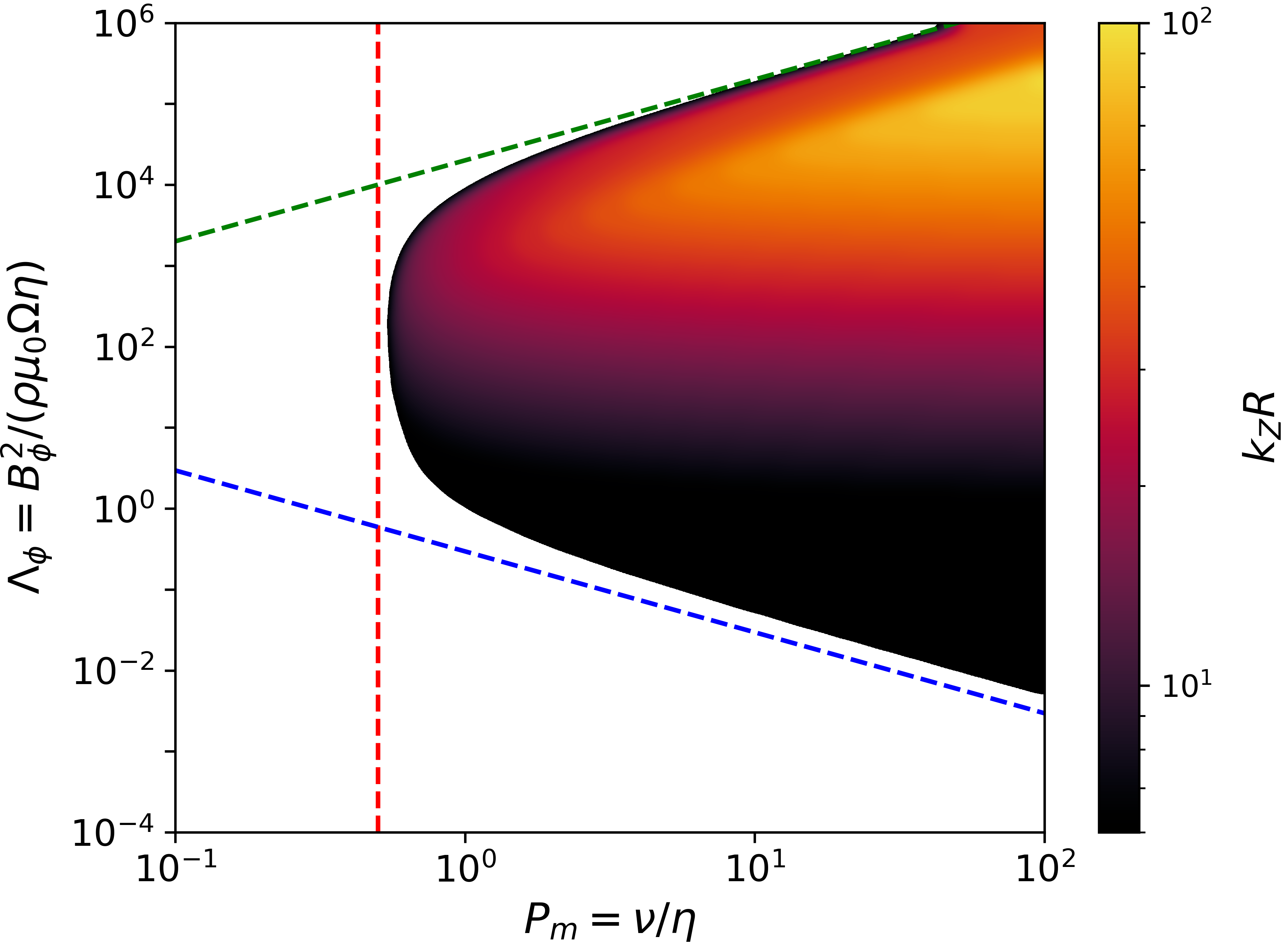}
  \caption{\textbf{Top panel:} Mode $k$ corresponding to the SMRI modes displayed in Fig.~\ref{expli} (top panel) \textbf{Bottom panel:} Mode $k$ corresponding to the AMRI modes displayed in Fig.~\ref{amri} (top panel).
  }
     \label{kexpli}
\end{figure}

A condition for the local approach to be valid is that the unstable modes must remain small compared to the system size and not larger than the typical domain over which the background parameters vary. This motivates us to derive reliable estimates of the characteristic wave-numbers for both the SMRI and the AMRI.
For the SMRI, the mode scale corresponds to the balance between magnetic tension, diffusion, and shear. In contrast, the AMRI is constrained by coupled diffusive frequencies that depend on both the magnetic field strength and the stratification. Since the azimuthal magnetic tension does not involve $k_Z$, the AMRI mode scale is governed by different physical effects than in the SMRI and, apart from secondary corrections, is essentially independent of the shear. We note that for both SMRI and AMRI, the most unstable mode has the lowest radial wavenumber of the values considered, $k_R = k_{R,min}$, and we  focus our study on the axial wavenumber $k_Z$.
In the case of the SMRI (see Fig.~\ref{kexpli}), again two regimes can be distinguished from the ratio of magnetic diffusivity and resistivity. The estimate given below Eq.~\ref{kSMRI} agrees well with that proposed by \citet{Sano1999} for protoplanetary disks, and with \citet{Petitdemange2008} (magnetostrophic regime) and reproduces the numerical values,
\begin{equation}
    k_{Z,\mathrm{SMRI}} R = \sqrt{\dfrac{2 \Lambda_Z \pram \vert \ross \vert }{(6+\Lambda_Z^2)\ekma}}.
\label{kSMRI}
\end{equation}
This estimate remains valid for more diffusive regimes (see Fig.~\ref{kexpli} top panel). 

For the AMRI modes, Fig.~\ref{kexpli} (bottom panel) shows the wavenumber $k_Z$ across different regimes. The coupled diffusive frequencies constrain the mode size as a function of magnetic field strength and stratification. The following estimate reproduces well the observed variation,
\begin{equation}
    k_{Z,\mathrm{AMRI}}R =  \dfrac{1}{2} \left[ \dfrac{m^2 \Lambda_\phi \pran \stra}{\ekma}  \right]^{1/4}. 
\label{kAMRI}
\end{equation}

\section{Equations governing the magnetic stabilisation of GSF-like instabilities}
\label{ApMGSF}
\subsection{Marginal stability equation}
When considering a purely axial magnetic field, the marginal stability condition reads
\label{sec:marginal}
\begin{equation}
\begin{aligned}
&\overbrace{\dfrac{\partial \ln \left(R^4 \Omega^2\right) }{\partial \ln R}}^{\text{GSF seed term}} 
+ \overbrace{2 \dfrac{\partial \ln \Omega}{\partial \ln R} \dfrac{k_Z^2 B_Z^2 }{k^4 \rho \mu_0 \eta^2 }}^{\text{SMRI seed term}} 
+ \overbrace{\dfrac{k_Z^2}{k^2} \left(\dfrac{B_Z^2}{\rho \mu_0 \eta \Omega} \right)^2}^{\text{Magnetic stress}} \\
+& \overbrace{\dfrac{2B_Z^2 \nu k^2}{\rho \mu_0 \eta \Omega^2}}^{\text{Mixed diffusion}} + \overbrace{\dfrac{\nu^2k^6}{\Omega^2 k_Z^2}}^{\text{Viscous}}
+ \overbrace{\dfrac{\nu}{\kappa} \left(\dfrac{N}{\Omega}\right)^2}^{\text{GSF stabilisation}} \\
+& \overbrace{\dfrac{\nu}{\kappa} \left(\dfrac{N}{\Omega}\right)^2 \dfrac{k_Z^2 B_Z^2}{k^4 \rho \mu_0 \eta \nu}}^{\text{Magneto-stratification}} 
 = 0.
\end{aligned}
\label{marginal_stability}
\end{equation}

Some contributions can be directly related to the three SMRI stability criteria introduced in Eq.~\ref{MRI_crit}. The SMRI seed term is balanced by:
\begin{itemize}
\item the magnetic stress, which yields the magnetic stress criterion (i),
\item the combined effect of the GSF seed and stabilizing terms, leading to the magnetic diffusion criterion (ii),
\item the magneto-stratification term, associated with the stratification criterion (iii).
\end{itemize}
 In the following, we combine the last two terms under the compact notation $(N/\Omega)^2_{\star} = (N/\Omega)^2 + 2(\kappa/\eta) \partial \text{ln} \Omega / \partial \text{ln}R$.  
 The notation for the last two terms is chosen analogously to the GSF stabilizing contributions. The viscous term is straightforward, while the mixed diffusion term introduces a coupling between magnetic strength (limited by diffusion) and viscosity, thereby connecting both dissipative processes.

Then we can deduce that the GSF modes are constrained by,
\begin{equation}
\begin{aligned}
\dfrac{\nu^2 k^4}{\Omega^2} 
\left| \dfrac{\partial \ln \left(R^4 \Omega^2\right) }{\partial \ln R} \right|^{-1} < &\dfrac{k_Z^2}{k^2} < \left( \dfrac{\rho \mu_0 \eta \Omega}{B_Z^2} \right)^2 
\left| \dfrac{\partial \ln \left(R^4 \Omega^2\right) }{\partial \ln R} \right|, \qquad \\
\dfrac{\rho \mu_0 \eta \Omega^2}{2 B_Z^2 \nu} \left| \dfrac{\partial \ln \left(R^4 \Omega^2\right) }{\partial \ln R} \right| > &k^2 > \dfrac{k_Z^2}{k^2} \dfrac{B_Z^2}{\rho \mu_0 \eta \kappa} 
\left( \dfrac{N}{\Omega} \right)_\star^2 
\left| \dfrac{\partial \ln \left(R^4 \Omega^2\right)}{\partial \ln R} \right|^{-1}.
\end{aligned}
\label{ineq}
\end{equation}

In a compact form these inequalities are expressed as\begin{equation}
k < k_{max}, \quad \alpha< \alpha_{max}, \quad 
k^2 < \mathcal{A}_1 \alpha, \quad \mathcal{A}_2 \alpha< k
,\end{equation}
with
\begin{equation}
\begin{aligned}
k_{max} &= \sqrt{\dfrac{2(\vert \ross \vert -1)}{\Lambda_Z \ekma}}, &\qquad \alpha_{max} &= \dfrac{2\sqrt{\vert \ross \vert -1}}{\Lambda_Z}, \\
\mathcal{A}_1 &= \dfrac{2\sqrt{\vert \ross \vert -1}}{\ekma}, &\qquad \mathcal{A}_2 &= \sqrt{\dfrac{\Lambda_Z \pran \stra_\star}{4 \ekma (\vert \ross \vert -1)}}.
\end{aligned}
\end{equation}

\subsection{MGSF asymptotic regimes}
 \label{asymp}
 
The minimisation of all stabilizing terms over the mode ranges $(k_Z,k)$ is not straightforward. Therefore, to quantify the additional stabilisation induced by an axial magnetic field, we distinguish two regimes. These two underlying minimisation problems arise because depending on the ordering of magnetic, viscous and stratification effects (as specified by Eq.~\ref{ineq}), different terms dominate the stabilisation.

$1.$ First we consider the contributions of viscous and mixed dissipative effects, together with magnetic stress.
From the marginal stability equation (Eq.~\ref{marginal_stability}), the boundary of the unstable parameter domain is defined by the mode $(k_Z,k)$ minimizing the stabilizing function $f(k_Z,k)$ that reads
\begin{equation}
f(k_Z,k) =  \dfrac{k_Z^2}{k^2} \Lambda^2 + 2 \Lambda \ekma k^2.
\end{equation}
Because $f$ increases monotonically with respect to $k_Z$, we use this relation to fix $k_Z$ and minimise $f$ with respect to $k$. Solving the resulting equation $\partial_{k} f = 0$ for $k^4$ gives the unique positive root, which we denote as $k_f$,
\begin{equation}
k_f^4 = \dfrac{\Lambda k_Z^2}{2 \ekma}.
\end{equation}
The corresponding minimum value of the function, $f$, is then, 
\begin{equation}
\mathcal{C}_1 =\text{min}_{(k_Z,k)} f = 2\sqrt{2}\Lambda^{3/2}\ekma^{1/2}k_{Z,min}.
\end{equation}

$2.$ In a second regime, we considered viscous diffusion and the interplay between magnetic and stratification effects. The stabilizing function is expressed as
\begin{equation}
g(k_Z,k) =  \dfrac{\ekma^2k^6}{k_Z^2} + \dfrac{k_Z^2 \Lambda \pran \stra_\star}{\ekma k^4}. 
\end{equation}
The two interior stationarity conditions $\partial g / \partial k_Z = 0$ and  $\partial g / \partial k = 0$ lead to incompatible relations between $k$ and $k_Z$, and therefore no interior extremum exists in the $(k_Z,k)$ domain. The minimum must lie on the boundary of the admissible region and we evaluate $g$ at $k_Z = k_{Z,min}$. Using the same method as before, the stabilizing contribution is minimised with respect to $k$ at
\begin{equation}
k_g^{10} = \dfrac{2\Lambda \pran \stra_\star k_Z^4}{3 \ekma^3},
\end{equation}
so that the minimum value of $g$ is
\begin{equation}
\begin{aligned}
\mathcal{C}_2 = \text{min}_{(k_Z,k)}g = \dfrac{5}{2}\left(\dfrac{2}{3}\right)^{3/5} \Big[\ekma k_{Z, min}^2 (\Lambda \pran \stra_\star )^{3}\Big]^{1/5}. 
\end{aligned}
\end{equation}

Finally, since $g_(k_Z, k_g)$ is strictly decreasing as $k_Z$ decreases, the global minimum of $g$ over the allowed domain is indeed reached at the lowest admissible vertical wavenumber $k_Z = k_{Z,min}$, which validates our assumption.
 
\subsection{MGSF stability criteria}
\label{mgsf_criteria}

In order to obtain a sufficient MGSF stability criterion, we consider the minimisation relation $\text{min}_{(k_Z,k)} (f+g) \geq  \text{min}_{(k_Z,k)} f + \text{min}_{(k_Z,k)} g$, combine the minimum values of $f$ and $g$ independently, and get

\begin{equation}
\begin{aligned}
\pran \stra  +  \mathcal{C}_1 + \mathcal{C}_2 >  4\left( \vert \ross \vert-1 \right).
\end{aligned}
\label{MGSFcrit0}
\end{equation}

To get a more strict criterion, we assume that the instability boundary is controlled by the two asymptotic branches identified above, and we evaluate the marginal stability $k_f$ and $k_g$. This time we consider the minimisation relation $\text{min}_{(k_Z=k_{Z,min},k \in [k_f,k_g])} (f+g) \geq  \text{min}_{(k_Z,k)} (f+g)$. By doing so we obtain the necessary MGSF stability criteria,

\begin{equation}
\begin{aligned}
& \pran \stra + \mathcal{C}_1 + \dfrac{1}{8} \mathcal{C}_1 + 2\,\pran \stra_\star
> 4\big(|\ross| - 1\big),\\
& 
   \pran \stra  + \mathcal{C}_2 + \frac{12\,\mathcal{C}_2^2}{25\,\pran \stra_\star}
   + \frac{18\,\mathcal{C}_2^3}{125\,\pran^2 \straquatre_\star} > 4\big(|\ross| - 1\big).
\end{aligned}
\label{MGSFnece}
\end{equation}

The relevance of the last criteria (Eq.~\ref{MGSFnece}) is illustrated in Fig.~\ref{GSFcri}, where the associated grey dashed line, representing the most restrictive necessary condition, closely follows the numerically determined boundary of the unstable parameter domain. Finally, because of the dependence of $C_2 \propto [(\nu/\kappa)(N/\Omega)^2]^{3/5}$, the standard GSF balancing term $(\nu/\kappa)(N/\Omega)^2$ ultimately dominates at large stratification numbers. Moreover, since $(N/\Omega)^2_\star$ represents the effective stratification arising from the SMRI balance (Eq.~\ref{MRI_crit},$iii$), it is negative in the SMRI parameter regime. Therefore, when estimating the minimal shear $\vert R_{o,min}\vert$ required for the onset of the MGSF instability, but not the SMRI, we obtain
\begin{equation}
 \vert R_{o,min}\vert \in \Big[1  + \pram + \dfrac{8\mathcal{C}_1}{32}, 1  + \pram + \dfrac{9\mathcal{C}_1}{32} \Big].
\end{equation}
In stellar radiative regions $1 \gg \pram$, so only $C_1$ needs to be retained, and this results applies throughout the entire MGSF parameter domain.

\begin{figure}
\centering
\includegraphics[width=9cm]{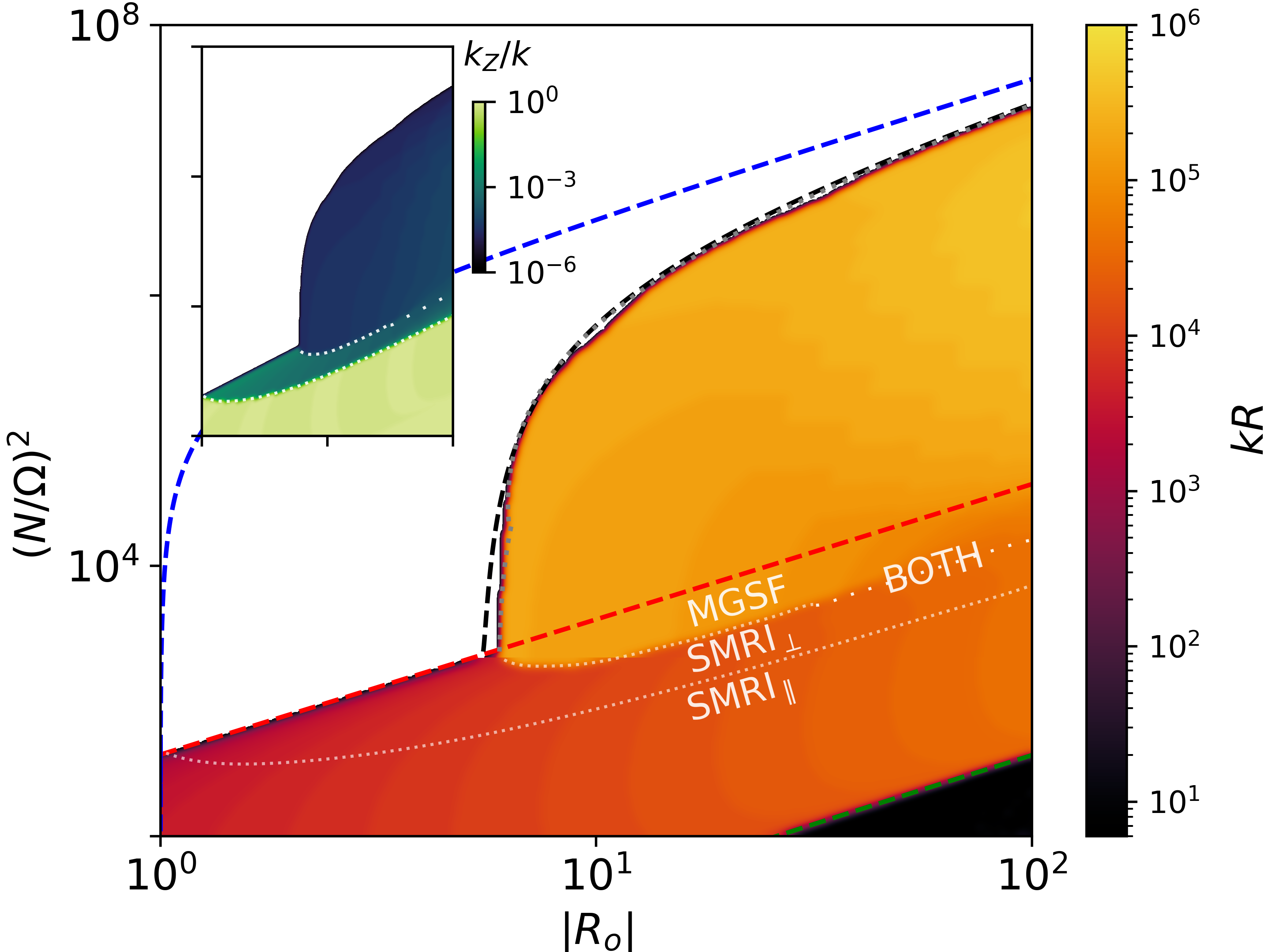} 
\caption{Mode wavenumber, $k$, and orientation, $k_Z/k$, for the instabilities found in the regime of Fig.~\ref{GSFcri}. Different instability domains are identified as follows: SMRI$_\perp$ modes share the growth rates of SMRI$_\parallel$ but are oriented nearly radially. As stratification increases, the SMRI branch is stabilised and MGSF modes dominate. These MGSF modes are clearly distinct from SMRI$_\perp$, exhibiting different growth rates, more horizontal orientation (lower $k_Z/k$), and smaller spatial scales (larger $k$).}
\label{kcri}
\end{figure}

\section{Method for global linear analysis}
\label{taylorcouette}
\subsection{Taylor-Couette framework and model assumptions}
\label{tc_profiles}
\begin{figure}
\centering
\includegraphics[width=4.5cm]{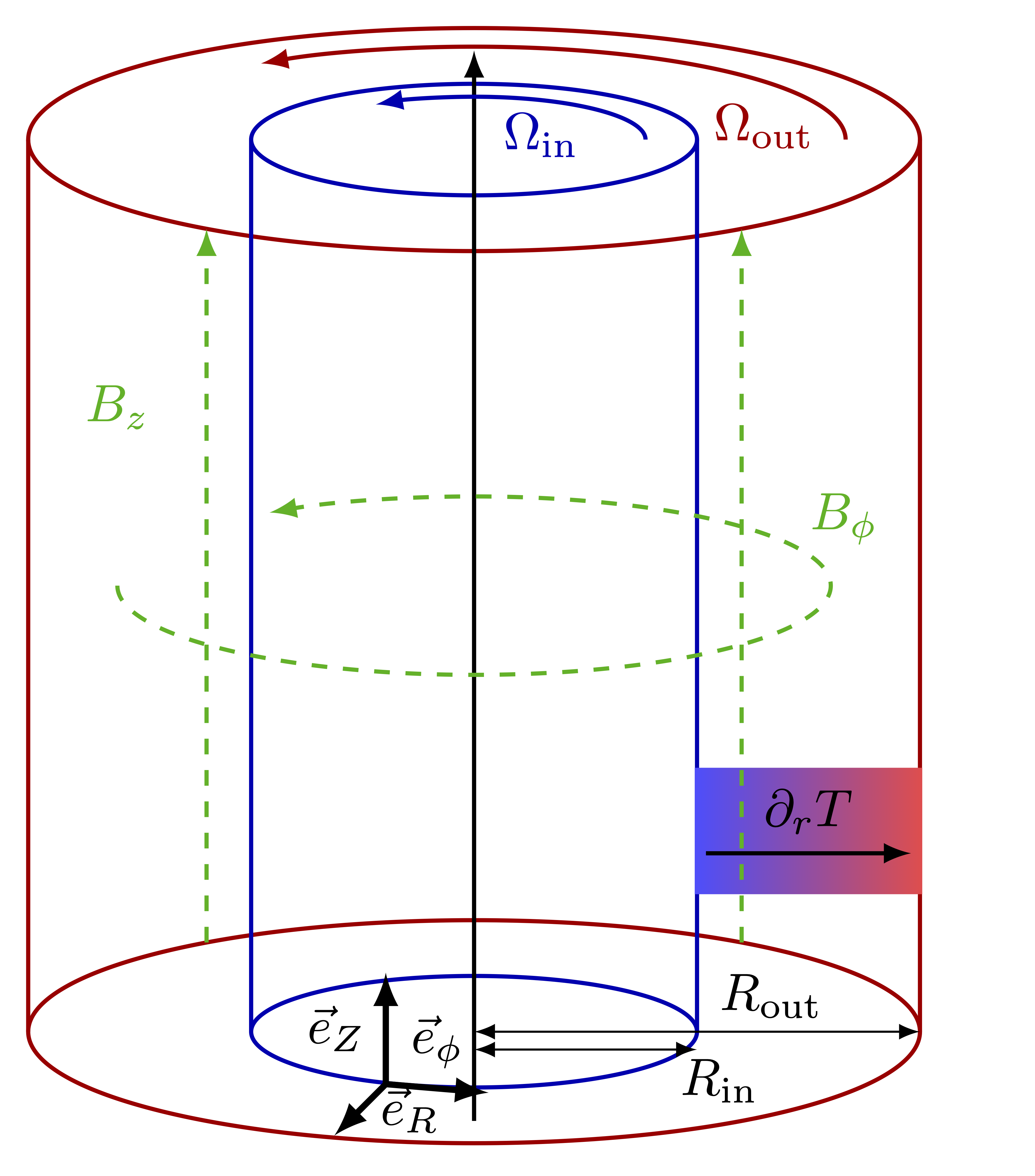} 
  \caption{Sketch of the system under study, showing a shear flow located between two differentially rotating cylinders, subject to a radial thermal gradient and immersed in a magnetic field with axial and azimuthal components.}
     \label{system_cyl}
\end{figure}

In the standard Taylor-Couette framework (see Fig.~\ref{system_cyl}), the background velocity and temperature fields satisfy the diffusive equilibrium conditions, namely $\nabla^2 \vec{U_0} = 0$ for the momentum equation and $\nabla^2 T_0 = 0$ for the thermal field. The basic profiles are $\Omega_{TC}(R) = A + B/R^2 $ and  $\mathrm{N}_{TC}^2 = \gamma g d_R T_0(R) = C/r$, with $\zeta = R_{in}/R_{out}$, $\xi = \Omega_{in}/\Omega_{out}$ so that $A = \Omega_{in} (\xi - \zeta^2)/(1-\zeta^2)$, $B = \Omega_{in} R_{in}^2 (1-\xi)/(1-\zeta^2)$, and $C= R\gamma g d_RT_0(R) = \gamma g (T_{out} - T_{in})/(\ln 1/\zeta)$.

Then, the perturbations $U^\prime, B^\prime$ are expressed in terms of poloidal
and toroidal components,
\begin{equation}
\begin{aligned}
\vec U' &= \vec \nabla \times (e \vec e_R) + \vec \nabla \times \vec \nabla \times (f \vec e_R), \\
\vec B' &= \vec \nabla \times (g \vec e_R) + \vec \nabla \times \vec \nabla \times (h \vec e_R),
\end{aligned}
\end{equation}
and together with the thermal perturbation, $T'$ (noted $j$ in analogy with \cite{Child2015}), we recover the system of five coupled linear equations. In the notation of \cite{Child2015}, the discretised system is expressed as
\begin{equation}
\begin{aligned}
&\gamma (C_2 e + C_3 f) + C_4 e + C_5 f = R_e (E_1 + F_1) + H_a^2 (G_1 + H_1), \\
&\gamma (C_3 e + C_4 f) + C_5 e + C_6 f = R_e (E_2 + F_2) + H_a^2 (G_2 + H_2 ) \color{blue} + R_e J_2, \color{black} \\
&P_m \gamma (C_1 g + C_2 h) + C_3 e + C_4 f = E_3 + F_3 + R_e P_m (G_3 + H_3), \\
&P_m \gamma (C_2 g + C_3 h) + C_4 e + C_5 f = E_4 + F_4 + R_e P_m (G_4 + H_4), \\&\color{blue} R_e P_r \gamma j = R_e P_r (N/\Omega)^2 F_5 + J_5.
\end{aligned}
\label{eq_glob}
\end{equation}
where $R_e = \Omega_{in} R_{in}^2/\nu$ is the Reynolds number, $H_a = B_0 R_{in}/\sqrt{\rho \mu \eta \nu}$ the Hartmann number (with $\vec{B} = B_0 \left(\delta \vec{e_Z} + \beta (R_{in}/R) \vec{e_\phi} \right)$), and we denote $J_2 = \Delta j, J_5 = -[i R_eP_r m + \Delta -(1/R)\partial_R -\partial^2_R]j, F_5 = \Delta f $ and $\Delta = k_Z^2 + m^2/R^2$. We expand the spatial structure of the perturbations onto a truncated spectral basis of $N_{rp}$ Chebyshev polynomials, and we find the eigenmodes using the same numerical method as described in Section~\ref{num_met}.

For the velocity, we adopt standard no-slip boundary conditions. For the magnetic field, we consider two types: perfectly conducting boundaries, defined by  $b_R = b_\phi/R + d_R b_\phi = 0$, and insulating boundaries, defined by  $\nabla \times \vec b \vert_{in/out} = 0$. In the insulating case, unstable modes can be confined in a thin region near the inner cylinder, with components that neither decay nor oscillate coherently. These correspond to wall modes and are not physically relevant, as coherent evolution of all field components is expected in realistic flows. Since our goal is to apply these results to stellar interiors, where no solid boundaries exist, the qualitative stability properties are not expected to depend strongly on the precise boundary choice. We therefore adopt perfectly conducting boundary conditions in this study.

Unless specified, in the dimensionless parameters used (Table~\ref{table}) to specify the regime studied, we adopt the maximum values of the angular velocity, Brunt–Väisälä frequency and azimuthal magnetic field profiles, and the typical length scale is $\bar R=\sqrt{R_{in}(R_{out}-R_{in})}$, following \cite{Rudiger2010}. For instance, we note $\Lambda_Z = B_Z^2 /(\rho \mu_0 \Omega_{in} \eta)$ and $\ekma = \nu/(\bar R ^2\Omega_{in})$.

\subsection{Other examples of mode localisation}
\label{loc}
\begin{figure}[ht]
\centering
\includegraphics[width=9cm]{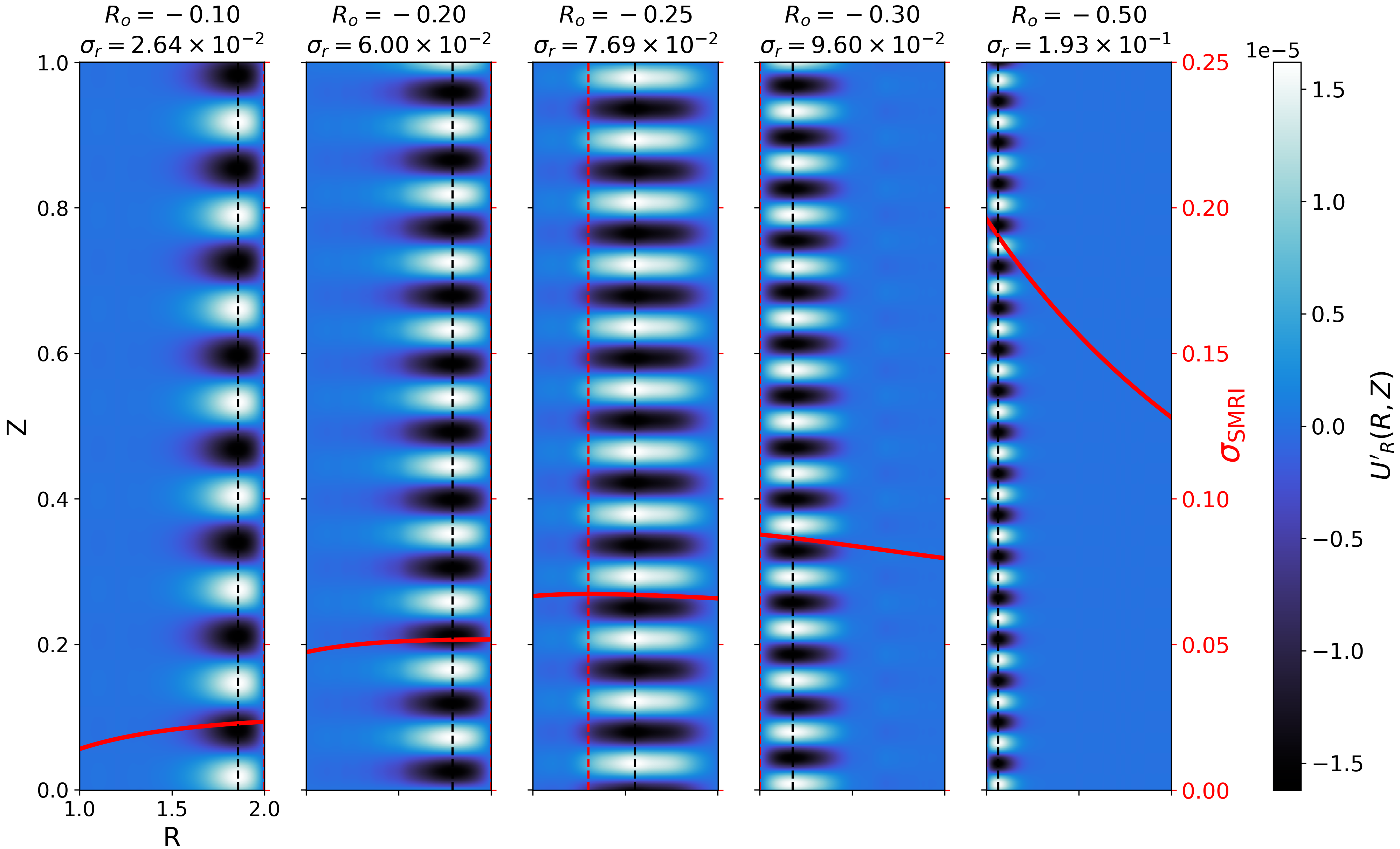}
\includegraphics[width=9cm]{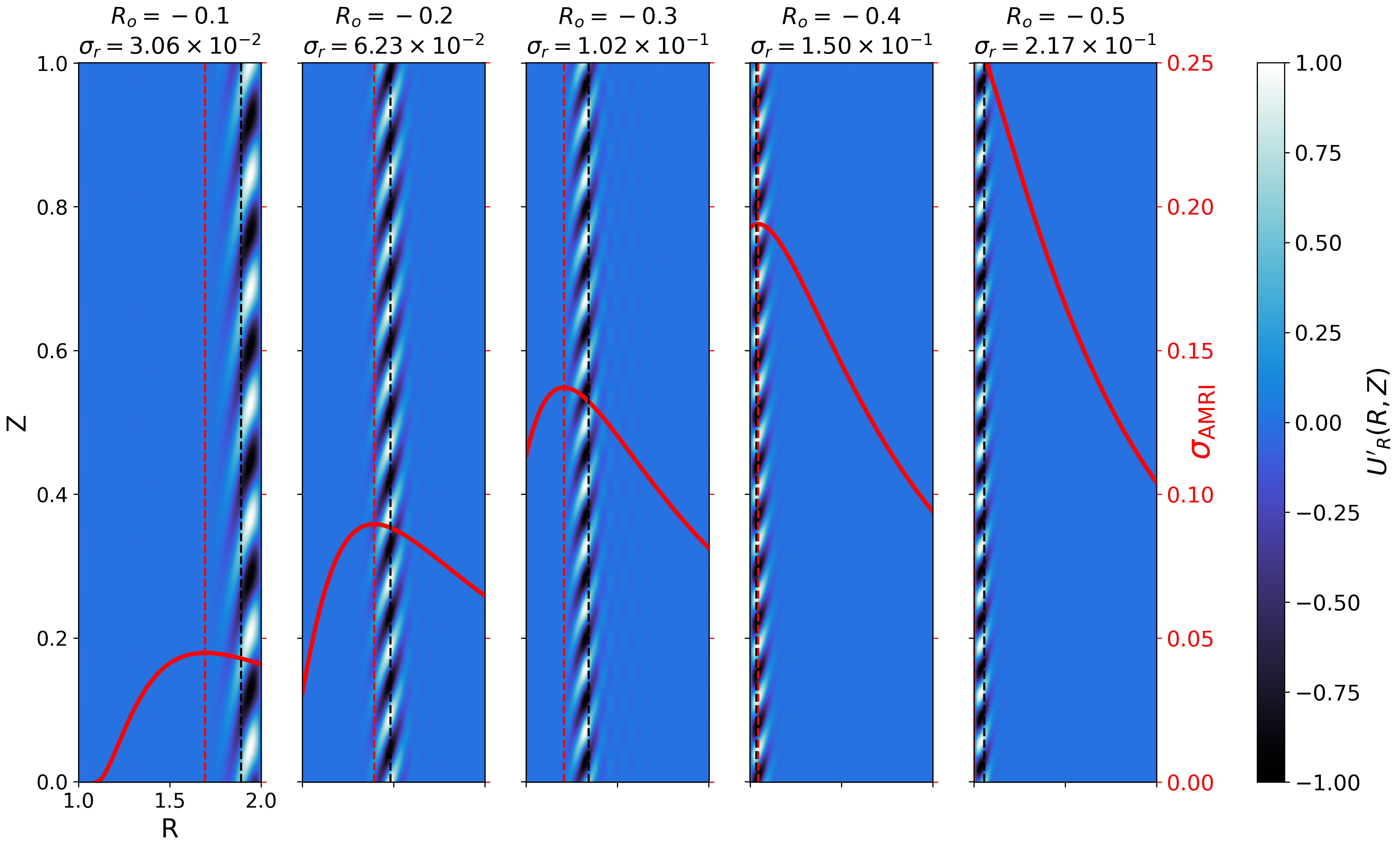}
  \caption{\textbf{Top panel:} Reconstructed radial velocity perturbations $U'_R(R,Z)$ of the most unstable SMRI eigenmode in the following regime, $\ekma=10^{-5}, \Lambda = 1.6, \pram=1, \pran=0.1, \beta =0$ for a radial stratification $\stra(R)=1.3/R^2$ and 5 different rotation profiles $\Omega(R) = \Omega_{in} r^{-2\ross}$ with a constant shear $\vert \ross \vert$ of $0.1, 0.2, 0.25, 0.3$ and $0.5$. The fastest growth rate computed is given below the shear value considered. The red overlayed curves correspond to the local estimate (Eq.~\ref{gr_model}). The other perturbations $U_\phi',U'_Z, B'_R, B'_\phi, B'_Z, T'$ are located in the same regions. \textbf{Bottom panel:} Same for AMRI eigenmode with $m=2$ and in the following regime, $\ekma=10^5, \Lambda_\phi = 2.5 \times 10^4, \pram = 2, \pran = 0.001, \stra (R)=100/R^2$ and 5 different rotation profiles $\Omega(R) = \Omega_{in} r^{-2\vert \ross \vert}$ with a constant shear $\vert \ross \vert$ of $0.1, 0.2, 0.25, 0.3$ and $0.5$.
          }
     \label{glob2_SMRI}
\end{figure}

In Fig.~\ref{glob2_SMRI} (top panel), we consider a steep Taylor-Couette thermal profile, $\mathrm{N}^2 = \mathrm{N}_{R=R_{in}}^2(R_{in}/R)^2$, together with an ad hoc rotation profile $\Omega(R)$ chosen such that the shear $\ross = (R/2\Omega)d_R \Omega$ remains constant. The parameters are $\ekma=10^{-5}, \Lambda = 1.6, \pram=1, \pran=0.1$. For several shear values, we display the radial velocity perturbations, with the local growth rate estimate along the radius shown in red. Discrepancies in the mode location are interpreted as arising from the nearly flat local growth rate profile, where non-local effects, such as curvature and boundary conditions, influence the position. We also note that the radial mode scale appears correlated with the sharpness of the local growth rate profile. The numerical growth rate is again in good agreement with the maxima reached by the analytical formula (Eq.~\ref{gr_model}) in the domain. Their ratio lie in $[0.98, 1.16]$, and as suggested by the local approach, the axial wavenumber scaling $k_{Z} \propto \sqrt{\vert \ross \vert}$ is recovered, see Table~\ref{tab:kglob_combined}.

\begin{table*}
\caption{Stability properties for the Star A $(r_b = 0.0146 R_\star, w_b = 0.0234 R_\star, \Omega_{in} = 3.17 \times 10^{-6}s^{-1})$ and the Star F $(r_b = 0.0092 R_\star, w_b = 0.0084 R_\star, \Omega_{in} = 1.03 \times 10^{-5}s^{-1})$ for different profiles $(r_d, w_d, B_{Z,0})$.}

\begin{center}
\begin{tabular}{l l l | l l l l l}
\hline\hline                                                                         \\[-7pt]
 $r_d$    &  $w_d$    & $B_{Z,0}$ & $\tau_r$ & $ k$ & SDI & $r_{loc}$ & $\Delta r$ \\
$[r_b]$ & $[w_b]$    & [Gauss]   &   & $[1/R_\star]$ & sub-kind & $[R_\star]$ &$[R_\star]$ \\
  \hline 
  \hline
  Star A & &  & & & & & \\
  \hline
  \hline
$[1;5]$ &  3 & $1 $ & -  & - & - & -  & - \\
 2 & 1 & 1 & -  & -  & -  & - & - \\
  3 & 1 & 1 & $9.85 \, \rm d$  & $7.40 \times 10^6$  & \textbf{G}  & 0.118 &  [0.111;0.126]\\
   &  & $10^3$ & $484 \, \rm yr$  & $7.28 \times 10^4$  & \textbf{M}  & 0.118 &  [0.113;0.123]\\
    & & $10^5$ & - & - &  -  & - & - \\
$3$ &  2 & $1$ & -  & - & - & -  & - \\
 5 & 1 & $1 $ & $24.1 \, \rm d$  & $1.47 \times 10^6$  & \textbf{G}  & 0.213 &  [0.200;0.232]\\
   & & $10^3$ & $1.41 \, \rm yr$ & $5.82 \times 10^5$ & \textbf{M} & 0.213 & [0.203;0.222]\\
  & & $10^5$ & $46 \, \rm kyr$ & $3.21 \times 10^3$ & \textbf{S}$_\perp$ & 0.213 & [0.203;0.222]\\
 5 & 2 & $1 $ & $7.04 \, \rm d$  & $8.30 \times 10^6$  & \textbf{G}  & 0.218 &  [0.202;0.239]\\
  &  & $10^3$ & $12.6 \, \rm yr$  & $4.12 \times 10^5$  & \textbf{M}  & 0.220 &  [0.202;0.239]\\
  & & $10^5$ & - & - & - & - & - \\
 \hline
linear &  - & $1$ & $15.7 \, \rm kyr$  & $4.05 \times 10^3$  & \textbf{S$_\parallel$}    & 0.58  &  [0.42,0.58]\\
linear &  - & $10^5$ & $68 \, \rm yr$  & $1.62 \times 10^4$  & \textbf{S$_\parallel$}    & 0.58  &  [0.42,0.58]\\
  \hline 
  \hline
  Star F & &  & & & & & \\
  \hline
  \hline
1 & $[1;6]$ & $1$     & -  & -  & -  & -  &  - \\
 1.25 &  3 & $1 $ & $4.5 \, \rm d$  & $5.87 \times 10^6$  & \textbf{G}     & 0.056  &  [0.044;0.064]\\
 &   & $10^5 $ & - & -  & - & -  & - \\
 1.5 &  3 & $1$     & $1.83 \, \rm d$  & $5.23 \times 10^6$  & S$_\parallel$ + \textbf{G}  & 0.074  &  [0.056;0.113]\\
  &   & $10^2$     & $12.1 \, \rm d$  & $1.31 \times 10^6$  & \textbf{S$_\parallel$ + M}    & 0.076  &  [0.056;0.113]\\
   &   & $10^5$     & $32 \, \rm kyr$  & $1.28 \times 10^3$  & \textbf{S$_\parallel$ + S$_\perp$ } & 0.075  &  [0.056;0.091]\\
2 & 3  &  1  & $20 \, \rm hr$ & $4.66 \times 10^6$ &  S$_\parallel$ + \textbf{G}      & 0.138  &  [0.094;0.189]\\
  &  & $10^{2}$ & $20.3 \, \rm hr$ & $3.70 \times 10^6$ & S$_\parallel$ + \textbf{G} & 0.138 & [0.094;0.189]\\
   &  & $10^{3}$ & $1.27 \, \rm d$ & $1.31 \times 10^6$ & \textbf{S$_\parallel$} + M & 0.138 & [0.094;0.189]\\
 &   & $10^5$ & $11.8 \, \rm yr$ & $1.62 \times 10^4$ &  \textbf{S$_\parallel$} + S$_\perp$      & 0.138   &  [0.094;0.175] \\
 &   5 & $10^5$ & $31.8 \, \rm yr$ & $1.29 \times 10^4$ &  \textbf{S$_\parallel$} + S$_\perp$      & 0.145   &  [0.081;0.199] \\
 \hline
 linear &  - & $1$ & $2.87 \, \rm d$  & $4.15 \times 10^6$  & S$_\parallel$ + \textbf{G}     & 0.317  &  [0.082;0.374]\\
linear &  - & $10^5$ & $353 \, \rm d$  & $5.15 \times 10^4$  & \textbf{S$_\parallel$}     & 0.364  &  [0.082;0.374]\\
 \hline\hline  
\end{tabular}
\label{table:resucesam}
\end{center}
\tablefoot{The column labelled as shear-driven instabilities (SDI) specifies the instability nature (G:GSF, M:MGSF, S$_{\parallel \text{ or } \perp}$ :SMRI$_{\parallel \text{ or } \perp}$). The delimiters $+$ specify the existence of different instability domains and the one that dominates is in bold. The column labelled as $r_{loc}$ specifies the expected main location of the instability and its maximal width, $\Delta r$.}
\end{table*}

\begin{table}[htbp]
  \caption{Axial wavenumber of the most unstable SMRI modes.}
    \label{tab:kglob_combined}
    \centering
    \renewcommand{\arraystretch}{1.2}
    \small
    \begin{tabular}{|c|c|c||c|c|c|}
        \hline
        \multicolumn{3}{|c||}{Regimes of Figure~\ref{glob2_SMRI}} & \multicolumn{3}{c|}{Regimes of Figure~\ref{glob_SMRI}} \\
        \multicolumn{3}{|c||}{varies $\vert \ross \vert$ (top panel)} & \multicolumn{3}{c|}{varies $\stra$} \\
        \hline
        $\vert \ross \vert$ & $k_{Z,glob}$ & $k_{Z,\mathrm{SMRI}}$
        & $\stra$ & $k_{Z,glob}$ & $k_{Z,\mathrm{SMRI}}$ \\
        \hline\hline
        0.1 & 48.9 & 55.5 & 1 & 25.1 & 33.5 \\
        \hline
        0.2 & 67.3 & 74.3 & 1.3 & 23.5 & 29.2 \\
        \hline
        0.25 & 73.3 & 88.0 & 1.5 & 21.0 & 27.1 \\
        \hline
        0.3 & 88.4 & 100.6 & 2 & 18.8 & 23.4 \\
        \hline
        0.5 & 110.4 & 130.0 & 2.5 & 17.9 & 20.8 \\
        \hline
    \end{tabular}
\tablefoot{We compare the mode wavenumbers obtained with a Taylor-Couette solver and the local approach for kinds of regime were studied, left columns varying $\vert \ross \vert$ constant in the domain, right columns varying the stratification factor $\stra$.}    
\end{table}

In Fig.~\ref{glob2_SMRI} (bottom panel), we perform a similar analysis for the AMRI, for the same kind of $\rm N^2(R)$ and $\Omega(R)$ profiles, with an azimuthal magnetic field and perturbations with azimuthal wavenumber $m=2$ and a different set of flow parameters, $\ekma =10^5, \Lambda_\phi = 2.5 \times 10^4, \pram = 2, \pran = 0.001, \stra=100$. We find a good correlation between the mode localisation and the maximum of the local growth rate profile, with comparable values. However, for parameter regimes closer to the stability boundary in the parameter space, the correlation becomes less robust, suggesting that curvature or boundary effects may play a significant role.

\subsection{Numerical stability of global MGSF modes}
\label{sec:mgsf_numstab}

In Fig.~\ref{GSF_glob}, the global modes grow more slowly than predicted by the local analytical formula (Eq.~\ref{estimMGSF}). We remark that the GSF and MGSF growth rates are sensitive to the radial resolution parameter $N_{rp}$ (here $N_{rp}=40$). Increasing $N_{rp}$ slightly raises the ratio between numerical and local values, suggesting that the preferred radial scale of GSF modes is very small and requires high resolution to be properly resolved. This behaviour is consistent with the local analysis of the MGSF (Fig.~\ref{sSMRI} bottom panel), involving fine spatial scales. At the same time, higher resolution enlarges the parameter range where spurious eigenvalues appear. For the regime displayed in Fig.~\ref{GSF_glob}, we made sure to achieve numerical convergence for the entire unstable parameter space.

\section{Instability properties for different models of subgiants and young RGB stars}
In Table~\ref{table:resucesam}, we summarise the instabilities and their properties for models with the different rotation and magnetic-field profiles described in Sect.~\ref{sec:global}. The large variability in the instability growth times indicates that stellar parameters shape instabilities of widely differing efficiencies, highlighting the need to include these effects in order to obtain realistic estimates of angular momentum transport throughout stellar evolution.

\label{cesamodels}

\end{appendix}
%
%
\end{document}